\documentclass[sigconf]{acmart}

\usepackage{booktabs} 

\setcopyright{rightsretained}

\usepackage{graphicx}
\usepackage{subfig}
\usepackage{float}
\usepackage{afterpage}
\usepackage{color}
\usepackage{url}
\usepackage{bbm}
\usepackage{epsfig}
\usepackage{epstopdf}
\allowdisplaybreaks

\newcommand{\eps}{\epsilon}
\newcommand{\abs}[1]{\left| #1 \right|}
\newcommand{\norm}[1]{\left\lVert #1 \right\rVert}

\newcommand{\X}{\mathcal{X}}

\renewcommand{\O}{\mathcal{O}}

\newcommand{\Q}{\mathcal{Q}}
\newcommand{\M}{\mathcal{M}}
\renewcommand{\S}{\mathcal{S}}
\renewcommand{\L}{\mathcal{L}}

\newcommand{\ED}{\text{ED}}
\newcommand{\HAM}{\text{HAM}}

\newcommand{\ecolis}{{\tt E.coli-small}}
\newcommand{\ecoli}{{\tt E.coli}}
\newcommand{\scere}{{\tt S.cerevisiae}}
\newcommand{\human}{{\tt Human}}

\newcommand{\mhap}{{\tt MHAP}}
\newcommand{\mini}{{\tt Minimap}}
\newcommand{\miniD}{{\tt Minimap-default}}
\newcommand{\miniA}{{\tt Minimap-alternative}}
\newcommand{\dalign}{{\tt DALIGNER}}
\newcommand{\blasr}{{\tt Blasr}}
\newcommand{\pac}{{\tt PacBio}}
\newcommand{\our}{{\tt SmoothQGram}}

\newcommand{\qG}{{$q$-gram}}
\newcommand{\sqG}{{smooth $q$-gram}}
\newcommand{\smG}{{smooth $m$-gram}}
\newcommand{\qGs}{{$q$-grams}}
\newcommand{\sqGs}{{smooth $q$-grams}}

\renewcommand{\paragraph}[1]{\medskip \noindent {\bf #1.}}

{\unskip\nobreak\hskip 1em plus 1fil\nobreak$\Box$
\parfillskip=0pt%
\endtrivlist}

\usepackage{algpseudocode}
\usepackage{algorithmicx}
\usepackage{algorithm}

\algnewcommand\algorithmicforeach{\textbf{for each}}
\algdef{S}[FOR]{ForEach}[1]{\algorithmicforeach\ #1\ \algorithmicdo}

\usepackage{multirow}
\usepackage{array}
\usepackage[mathscr]{eucal}
\usepackage{chngpage}

\copyrightyear{2018}
\acmYear{2018}
\setcopyright{acmlicensed}
\acmConference{}{}{}
\acmPrice{15.00}
\acmDOI{}
\acmISBN{}

\begin{document}
\title{Smooth $q$-Gram, and Its Applications to Detection of Overlaps among Long, Error-Prone Sequencing Reads}

\author{Haoyu Zhang}
\affiliation{%
  \institution{Indiana University Bloomington}
  \city{Bloomington} 
  \state{IN} 
  \postcode{47408}
  \country{USA}
}
\email{hz30@umail.iu.edu}

\author{Qin Zhang}
\affiliation{%
  \institution{Indiana University Bloomington}
  \city{Bloomington} 
  \state{IN} 
  \postcode{47408}
  \country{USA}
}
\email{qzhangcs@indiana.edu}

\author{Haixu Tang}
\affiliation{%
  \institution{Indiana University Bloomington}
  \city{Bloomington} 
  \state{IN} 
  \postcode{47408}
  \country{USA}
}
\email{hatang@indiana.edu}

\begin{abstract}
We propose {\em \sqG}, the first variant of $q$-gram that captures \qG\ pair within a small edit distance.  We apply \sqG\ to the problem of detecting overlapping pairs of error-prone reads produced by single molecule real time sequencing (SMRT), which is the first and most critical step of the {\em de novo} fragment assembly of SMRT reads.  We have implemented and tested our algorithm on a set of real world benchmarks. Our empirical results demonstrated the significant superiority of our algorithm over the existing \qG\ based algorithms in accuracy.
\end{abstract}

\maketitle

\section{Introduction}
\label{sec:intro}

$Q$-gram, also called $n$-gram, $k$-mer/shingle, has been used extensively in the areas of bioinformatics~\cite{AS90, PTW2001,BKC15, L16, M14}, databases~\cite{XWL08, WLF12, QWL11}, natural language processing \cite{M99}, etc.  In particular, \qG\ was used to construct the {\em de Bruijn graph} \cite{PTW2001,Compeau2011}, a data structure commonly exploited for fragment assembly in genome sequencing, especially for short reads obtained using next-generation sequencing (NGS) technologies ~\cite{Miller2010}. Another important application of \qG\ in bioinformatics is in {\em sequence alignment}, which aims to detect highly similar regions between long strings (e.g., genomic sequences). Following the {\em seed-extension} approach, many sequence alignment algorithms (including the popular BLAST \cite{AS90} and more recent algorithms ~\cite{Brudno2003,Schwartz2003,Kurtz2004}) first search for \qG\ matches (i.e., {\em seeds}) between each pair of input strings, and then extend these matches into full-length alignment by using dynamic programming algorithms. Recently, this approach was adopted for detecting {\em overlaps} between {\em long, error-prone reads} \cite{BKC15, L16, M14} generated by single molecule (also called the third generation) sequencing technologies, including the {\em single molecule real time sequencing (SMRT)} \cite{Roberts2013} and the {\em MinION sequencers} \cite{Mikheyev2014}. Comparing with the NGS reads, the single molecule technologies generate {\em reads} much longer and more error-prone. As a result, two overlapping reads contain highly similar but not identical substrings (with a relatively small edit distance\footnote{The edit distance between two strings $x$ and $y$ is defined to be the minimum number of letter insertions, deletions and substitutions needed to transfer $x$ to $y$.} due to sequencing errors), which should be addressed by an overlap detection algorithm.

A straightforward application of the {\em seed-extension} approach to overlap detection may be hurdled by an inherent limitation: two strings sharing a highly similar substring may share only a small number of, or even zero, matched \qG\ pairs ({\em seeds}), due to the pattern of sequencing errors within the shared substring. Consequently, a seed-extension algorithm may fail to detect such overlaps because of the lack of seeds between the reads.  
Let us illustrate this point by an example.  Consider the following two input strings:
\begin{eqnarray*}
 00000 \ 00000 \ 00000 \ 00000 \ 00000 \ 0000, & \text{and} \\
 00000 \ 00001 \ 00000 \ 00001 \ 00000 \ 0000, &
\end{eqnarray*}
Their edit distance is $2$, however, they share no matched $10$-gram pairs (seeds).

To address this issue, in this paper, we propose a variant of \qG\ called the {\em \sqG}, using which we can identify not only those exactly matched  \qG\ pairs (with certainty), but also those \qG\ pairs that have small edit distances (each with a high probability).  Our \sqG\ construction is based on a recent advance in metric embedding \cite{CGK16} that maps a string from the edit distance space to the Hamming distance space while (approximately) preserving the distance; we will illustrate the details of this embedding in Section~\ref{sec:CGK}.  For the example mentioned above, our \sqG\ based approach can, with a very high probability, find most pairs of \qGs\ of the two input strings whose edit distances are at most $1$.

\paragraph{Application in SMRT data}
We applied the \sqG\ to the overlap detection among sequencing reads produced by SMRT, which is the first and most critical step of the {\em de novo} fragment assembly of SMRT reads. Notably, SMRT sequencers generate reads of 1,000-100,000 bps long with 12-18\% sequencing errors (including most insertions/deletions and some substitutions); in comparison, Illumina sequencers (a common NGS platform) generate reads of 100-300 bps long with $<1\%$ errors .  We have evaluated our approach using real-world SMRT datasets.

We formalize the {\em overlap detection problem} as follows. Given a collection of strings $\X = \{x_1, \ldots, x_n\}$, the goal is to output all overlapping string pairs $\{(x, y)\ |\ x, y \in \X\}$ and their shared substrings $x_{sub}$ and $y_{sub}$, such that the lengths of the substrings are above a threshold $\Gamma$, and their edit distance is below a threshold $\theta$. 

\begin{figure}
\centering
\includegraphics[height = 1.6in]{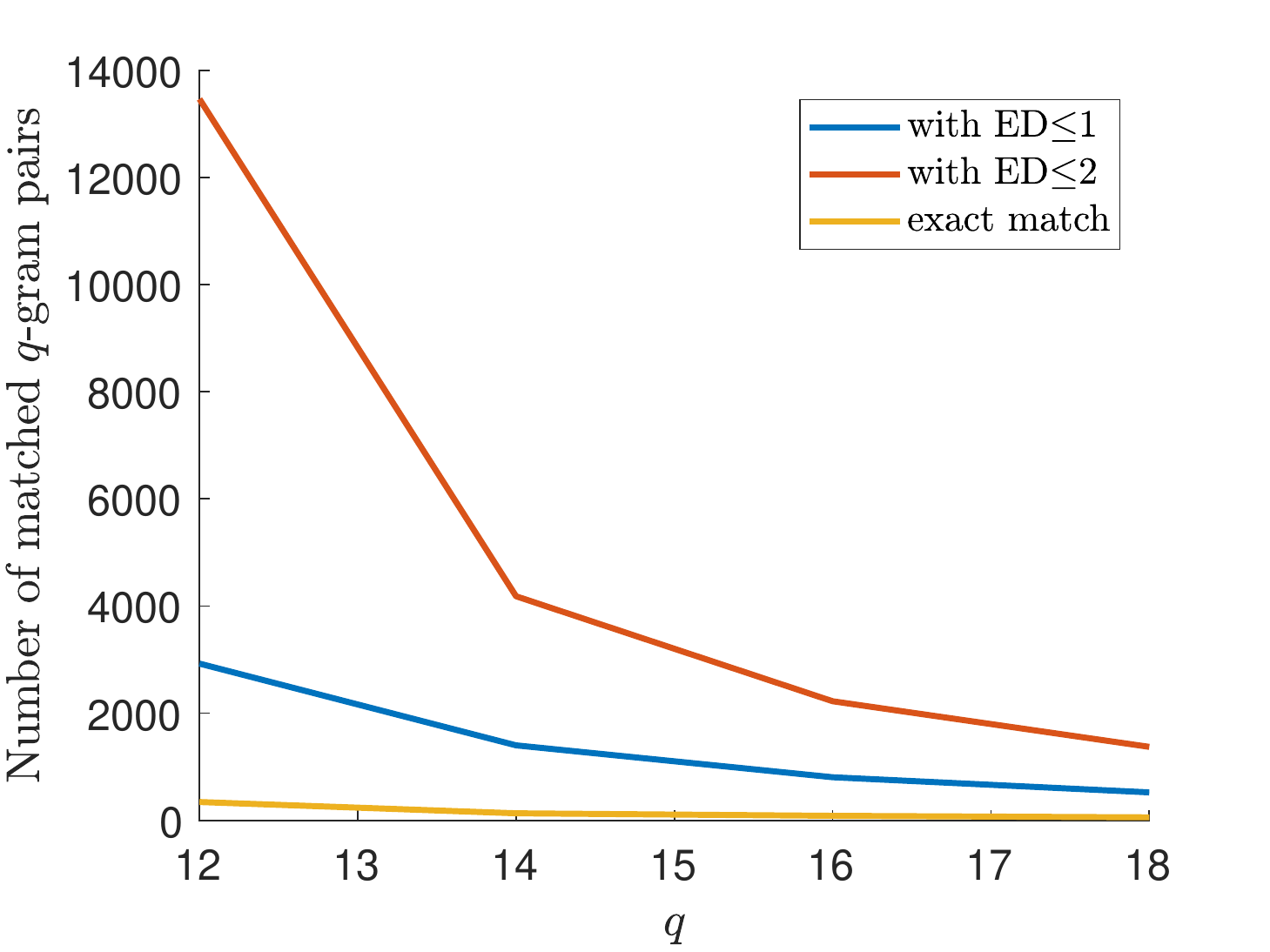}
\caption{Average number of matched $q$-gram pairs for $50$ overlapping read pairs in the \ecoli\ dataset.}
\label{fig:nummatch}
\end{figure}

For long, error-prone reads produced by SMRT, finding a good number of exactly matched \qGs\ between reads could be difficult (or just impossible). Thus the overlap detection problem becomes challenging for conventional ``seed-extension" approaches.  For example, in Figure~\ref{fig:nummatch} we plotted the average number of matched \qGs\ between $50$ overlapping SMRT reads sampled from a real word dataset \ecoli\ under different matching thresholds (edit distance threshold being $0$, $1$, and $2$, respectively).  We can see that when $q = 12$, the number of \qG\ pairs with edit distance (ED) no more than $2$ is $39.2$ times of that of exactly matched \qG\ pairs.  Obviously, for a pair of reads, with more matched  \qGs\ (seeds) detected, the sensitivity increases for detecting putative overlaps between error-prone reads. Therefore, the \sqG\ approach proposed in this paper can outperform the existing \qG\ based ``seed-extension" approaches. Indeed, our evaluation showed that, for the overlap detection in the real-world datasets that we have tested, the \sqG\ based algorithm {\em always} achieved $F_1$ scores (i.e., the harmonic average of the precision and recall) above $0.9$, while the $F_1$ score achieved by the best \qG\ based algorithm can be as low as $0.77$.

\paragraph{Our Contribution} We summarize our contribution below.
\begin{enumerate}
\item We proposed \sqG, the first variant of \qG\ that captures \qG\ pair within a small edit distance.

\item We applied \sqG\ to the problem of detecting overlapping pairs of error-prone reads produced by single molecule sequencing technologies, such as SMRT.

\item We implemented our \sqG\ based algorithm and tested it on a set of real world benchmarks. Our empirical results demonstrated the significant superiority of our algorithm over the existing \qG\ based algorithms in precision, recall and $F_1$ scores.
\end{enumerate}

\paragraph{Related Work}
Since it is proposed recently, the problem of detecting overlaps among long, error-prone reads from SMRT has drawn a significant attention in bioinformatics~\cite{BKC15, L16, M14}.  All existing overlap detection algorithms follow the ``seed-extension" approach, in which the seeds are defined based on $q$-grams. 

The only line of work, as far as we have concerned, that has a similar spirit as ours is the {\em gapped \qG} \cite{BK01,BK02,BK03} (also referred to as the {\em spaced seeds} in bioinformatics applications \cite{Ma2002,Keich2004}).  The idea of gapped \qG\ is to take substrings of each string of a specific pattern.  For example, the gapped $3$-grams of the string ``ACGTACGT'' with pattern ``XX-X'' are \{ACT, CGA, GTC, TAG, ACT\}. That is, instead of taking the contiguous substrings as that in the traditional \qG\ approach, the gapped \qG\  breaks the adjacency dependencies between the characters.  Now if we are allowed to choose multiple gapped \qG\ patterns, then one will need more edits to make all gapped \qGs\ between two strings mismatched. However, the optimal pattern of gapped \qG\ is difficult to find: it needs an exhaustive search on all possible patterns, and the running time for the search has an exponential dependency on length of the pattern \cite{Keich2004}. 
This might be the reason why there is no previous work applying gapped \qG\ to solve the overlap detection problem for SMRT data. In contrast, our \sqGs\ are systematically generated, and always have the same theoretical guarantees on all datasets.



\section{Smooth \lowercase{$q$}-Gram}
\label{sec:q-gram}

As mentioned, the major innovation of this paper is to replace the standard \qG\ based approach for overlap detection  with the \sqG\ based approach. The advantage of \sqG\ is that it tolerates a small edit distance between matched \qGs\ and is thus able to identify similar strings at higher sensitivity. In this section, we discuss the details of \sqG\ construction and discuss its properties. 

We will use $m$ to denote the length of a \sqG, and $\kappa$ to denote the length of a \qG\ after CGK-embedding.

%

\subsection{The CGK-Embedding}
\label{sec:CGK}

The key tool that we will use in our construction of \sqG\ is the CGK-embedding, which convert a string $s \in \Sigma^q$ to $s' = \Sigma^\kappa$ for a value $\kappa$ using a random string $R_1$, where $\Sigma$ is the alphabet (for nucleotides, $\Sigma = \{A, C, G, T\}$).  

More precisely, let $j = 1, 2, \ldots, \kappa$ denote the time steps of the embedding.  We also maintain a pointer $i$ to the string $s$, initialized to be $i = 1$.  At each step $j$, we first copy $s[i]$ to $s'[j]$, and set $j \gets j + 1$.  We then determine whether we should increment $i$ or not.   We sort characters in $\Sigma$ in an arbitrary but fixed order. For a character $\sigma \in \Sigma$, let $\text{Index}(\sigma)$ denote the index of $\sigma$ in this order. We set
$$i \gets i + R_1[j \cdot \abs{\Sigma} - \text{Index}(s[i]) + 1].$$  
When $i$ reaches $q + 1$ while $j < \kappa$, we simply pad $\kappa - j$ copies of `$\perp$' to $s'$ to make its length equal to $\kappa$, where $\perp \not\in \Sigma$ is an arbitrary character.

Denote the CGK-embedding as a function $CGK(\cdot, R_1)$ for a fixed string (sampled randomly from $\{0,1\}^{\kappa \abs{\Sigma}}$).  Given $s, t \in \Sigma^q$, let $s' = CGK(s, R_1)$ and $t' = CGK(t, R_1)$. It has been shown in \cite{CGK16} that for any $\kappa \ge 2q + c \sqrt{q}$ for some large enough constant $c$, we have with probability $0.999$ that 
\begin{equation*}
\label{eq:a-1}
\ED(s, t) \le \text{HAM}(s', t') \le O\left((\ED(s, t))^2\right),
\end{equation*}
where $\ED(\cdot, \cdot)$ and $\text{HAM}(\cdot, \cdot)$ denote the edit distance and the Hamming distance respectively.

It is easy to see that after the CGK-embedding, \qGs\ with small edit distance will likely have small Hamming distance, and those with large edit distance will likely have large Hamming distance.  In particular, if $s = t$, then we have $s' = t'$ with certainty.

The CGK-embedding has recently been used for sketching edit distance~\cite{BZ16} and performing edit similarity joins~\cite{ZZ17}.

\subsection{From $q$-Gram to Smooth $q$-Gram}
\label{sec:smooth}

We show how to construct a \sqG\ from a standard \qG\ using random string $R_2$.  For convenience we will write ``\sqG'' instead of ``\smG'' although the resulting \sqG\ will have length $m$.  Our algorithm is very simple.  Given a \qG\ $s$, we first perform the CGK-embedding on $s$ to get a string $s'$ of length $\kappa$, and then construct a substring $\bar{s}$ of length $m$ by picking the coordinates $i$ in $s'$ where $R_2[i] = 1$. The algorithm is depicted in Algorithm~\ref{alg:smooth-q-gram}.

\begin{algorithm}[t]
\caption{Generate-Smooth-$q$-Gram($s, R_1, R_2$)}
\label{alg:smooth-q-gram}
\begin{algorithmic}[1]
\Require $s$: $q$-gram $s\in \Sigma^q$;

$R_1$: random string from $\{0,1\}^{\kappa \abs{\Sigma}}$;

$R_2$: random string from $\{0,1\}^\kappa$ under the constraint that there are $m$ $1$-bit;

\Ensure  $\bar{s}$: smooth $q$-gram of $s$ of size $m$
\State $s' \leftarrow $ CGK($s$, $R_1$)  \label{ln:sig1}
\State $\bar{s}$ is generated by removing coordinates $i$ in $s'$ s.t. $R_2[i] = 0$
\State \Return $\bar{s}$
\end{algorithmic}
\end{algorithm}

The motivation of introducing \sqG\ is that we hope that the corresponding \sqGs\ of two \qGs\ $s$ and $t$ for which $\ED(s, t)$ is small, can be identical with a good probability.  More precisely, let $k = \ED(s, t)$, and let $s' = CGK(s, R_1)$ and $t' = CGK(t, R_1)$.  By the property of the CGK-embedding, we know that $\HAM(s', t') \le k^2$.  Let $d = \HAM(s', t')$. If we randomly sample without replacement $m$ bits from two $\kappa$-bit strings $s'$ and $t'$ at the same indices, the probability that all the sampled bits are the same is
\begin{eqnarray}
&&\frac{\kappa - d}{\kappa} \times \frac{\kappa - d - 1}{\kappa - 1} \times \cdots \times \frac{\kappa - d - (m - 1)}{\kappa - (m - 1)} \nonumber \\
&=& \frac{(\kappa - m) \times \cdots \times (\kappa - (d-1) - m)}{\kappa \times (\kappa - 1) \times \cdots (\kappa - (d-1))}. \label{eq:a-1}
\end{eqnarray}
In our experiments we typically choose $m = \kappa/c$ for a constant $c$, and we are only interested in $d$ being at most $4$.  In this case we can approximate (\ref{eq:a-1}) as $\left((c-1)/c\right)^d$ for some constant $c$.  In other words, for a non-trivial fraction of pairs of \qG, their corresponding \sqG\ will be matched.  Finally, we note that when $s = t$, with fixed $R_1$ and $R_2$ we must have $\bar{s} = \bar{t}$ with certainty.

We note that our construction of \sqG\ is very different from just a subsampling of the original \qGs.  Indeed, given two \qGs\ $s$ and $t$ where $t$ is obtained by a cyclic shift of $s$ by one coordinate (that is, we move the first coordinate of $s$ to the end of $s$, and $\ED(s,t) = 2$), if we just sample say a constant fraction of coordinates from $s$ and $t$ using common randomness, getting $\bar{s}$ and $\bar{t}$, then $\bar{s}$ and $\bar{t}$ will be different with very high probability.

As mentioned in the introduction, if we are able to match near-identical \qGs\ (under edit distance), then we are able to catch similar pairs of strings which will otherwise be missed by standard \qG\ approaches. In this way we can significantly improve the recall of the algorithm.  Of course, by allowing approximate matching we may also increase the number of false positives, that is, dissimilar pairs of strings may have many identical \sqGs, and will thus be considered as similar pairs. To maintain a good precision we may need to perform a verification step on the candidate pairs of similar strings, which will increase the running time.  Therefore, for a particular application, one needs to select a good tradeoff between the accuracy improvement and the extra running time cost.

We also comment that we can further enhance the precision by performing multiple CGK-embeddings (say, $d$ times), and/or multiple subsamplings (say, $z$ times), so that for each \qG\ we will create $d \times z$ \sqGs.  However, these operations will increase the number of false positives as well, and consequently  the running time.  In our experiments in Section~\ref{sec:exp} we have computed the number of matching \qGs\ on various datasets when varying the number of CGK-embeddings and subsamplings.  But for our application of detecting overlapping error-prone sequencing reads, we have noticed that a single run of CGK-embedding and subsampling already gives satisfactory accuracy.

\section{Applications to Overlap Detection among Long, Error-Prone Sequencing Reads}
\label{sec:algo}

In this section we show how to use \sqG\ to solve the overlap detection problem for long, error-prone sequence reads.  We approach the problem in two steps.  In the first step (Section~\ref{sec:candidate}), we show how to use \sqG\ to detect putative pairs of overlapping strings. And then for each of such pairs, we design an efficient verification procedure to reduce the number of false positives (Section \ref{sec:verification}).

In Table~\ref{tab:para-2} we have listed a set of global parameters/notations that will be used in our algorithms.  Let $[n]$ denote the set $\{1, 2, \ldots, n\}$.

\begin{table}[t]
\centering
\begin{tabular}{|c|c|} 
\hline
$m$ & length of \sqG \\
\hline
$\kappa$ & length of \qG\ after CGK-embedding \\
\hline
$\alpha$ & signature selection rate \\
\hline
$\eta$ & frequency filtering threshold \\
\hline
$K$ & edit distance threshold \\ 
\hline
$C$ & threshold for \#matched signatures \\
\hline
$L$ & targeting overlap length\\ 
\hline
$\epsilon$ & error tolerance rate\\
\hline
$\Pi$ & $\Pi : \Sigma^m \to (0, 1)$ a random hash function \\
\hline
$R_c$ & random string from $\{0,1\}^{\kappa \abs{\Sigma}}$ \\
\hline
$R_s$ & random string from $\{x \in \{0,1\}^\kappa\ |\ \norm{x}_1 = m\}$  \\
\hline
\end{tabular}
\caption{List of Global Parameters}
\label{tab:para-2}
\end{table}

\subsection{Detecting Putative Pairs of Overlapping Strings}
\label{sec:candidate}

Our algorithm for detecting overlapping pairs of strings is presented in Algorithm~\ref{alg:overlap}.

We will use the following data structure to store useful information of a \qG.
\begin{definition}[\qG\ signature]
\label{def:signature}
Let $\delta(s, t, r, i, p)$ be a signature for a \qG; the parameters are interpreted as follows: 
\begin{itemize}
\item[--] $s$ is the \qG; 

\item[--] $t \gets \text{Generate-Smooth-\qG}(s, R_c, R_s)$

\item[--] $r \gets \Pi(t)$, which can be seen as a hash rank of $t$; 

\item[--] $i, p$ denote that $s$ is taken from the $i$-th input string $x_i$ from the position $p$, that is, $s \gets x_i[p,p + q - 1]$.  
\end{itemize}
\end{definition}
It is easy to see that $t$ and $r$ are fully determined by $s$ given the randomness $R_s, R_c$ and $\Pi$, but for convenience we still include them as parameters in the definition of the signature.

\begin{algorithm}[t]
\caption{Find-overlapping-Strings($\X$)}
\label{alg:overlap}
\begin{algorithmic}[1]
\Require $\X = \{x_1, \ldots, x_n\}$: set of input strings; 

%
%
%
%
%

\Ensure  $\O \gets$ \{overlapping pair $(x_i, x_j)$ and their shared substrings $x_i[p_i^s, p_i^e]$ and $x_j[p_j^s, p_j^e] \}$

\State Initialize an empty table $D$  \label{ln:a-1}

\ForEach{$i \in [n]$} 
	\State $\Delta_i  \leftarrow \emptyset$
	\ForEach{$p \in [\abs{x_i} - q + 1]$}
		\State $s \leftarrow x_i[p, p + q - 1]$
		\State $t \leftarrow$ Generate-Smooth-$q$-Gram($s, R_c, R_s$)
		\State $r \leftarrow \Pi(t)$  
		\State $\delta \leftarrow (s, t, r, i, p)$
		\State $\Delta_i  \leftarrow \Delta_i  \cup \delta$
	\EndFor
\EndFor \label{ln:a-2}

\State Count for all $t$ the number $c_t$ of signatures in the form of $(\cdot, t, \cdot, \cdot, \cdot)$ in $\bigcup_{i \in [n]} \Delta_i$, and remove all $(s, t, r, i, p)$ in $\Delta_i$ with $c_t \ge \eta \sum_t c_t$ for all $i \in [n]$ \label{ln:a-21}

\ForEach{$i \in [n]$} \label{ln:a-3} 
	\State Construct $\Delta_i'$ from  $\Delta_i$ by keeping signatures in $\Delta_i$ with the smallest $\alpha |x_i|$ of hash ranks $r$. \label{ln:a-31} 
	\State $\L_i  \leftarrow \emptyset$
	\ForEach{$\delta$ in $\Delta_i'$}
		\State $\L_i \leftarrow \L_i \cup$ Search-Similar-$q$-Grams($\delta$, $D$)
	\EndFor
	\ForEach{$j < i$}
		\State $\M_{ij} \leftarrow \{(u, v)\ |\ (j, u, v) \in \L_i\}$  \label{ln:a-32} 
	\EndFor
\EndFor \label{ln:a-4}

\ForEach {$\abs{\M_{ij}} \ge C$} \label{ln:a-5} 
	\State  $(o, pos) \leftarrow$ Verify($x_i$, $x_j$, $\M_{ij}$)  \label{ln:a-51}
	\If{$(o, pos) \neq null$}
	\State $(p_i^s, p_i^e, p_j^s, p_j^e) \leftarrow$ Find-Shared-Substrings($x_i$, $x_j$, $o$, $pos$, $\Delta_i$, $\Delta_j$)
	\State $\O \leftarrow  \O \cup \left(x_i, x_j, [p_i^s, p_i^e], [p_j^s, p_j^e]\right)$
	\EndIf
\EndFor \label{ln:a-6}
\State \Return $\O$
\end{algorithmic}
\end{algorithm}

We now describe Algorithm~\ref{alg:overlap} in words.  The algorithm can be divided into three stages.  The first stage (Line~\ref{ln:a-1} -\ref{ln:a-2}) is the initialization: for each input string $x_i$, and for each of its \qG, we generate the corresponding \qG\ signature.  In the second stage (Line~\ref{ln:a-21} - \ref{ln:a-4}) we try to find a set of candidate overlapping pairs of  input strings.  We will explain how this works in the rest of this section.  The last stage (Line~\ref{ln:a-5}-\ref{ln:a-6}) is a verification step and will be illustrated in Section~\ref{sec:verification}.

In the second stage of the algorithm, the first step is to filter out those \sqGs\ whose frequency is above a certain threshold (Line~\ref{ln:a-21}).  This is a common practice, and has been used in a number of previous algorithms, such as  \mhap\cite{BKC15}, \mini\cite{L16}, \dalign\cite{M14}.  The motivation of this pruning step is that frequent \sqGs\ often correspond to frequent \qGs, which do not carry much important features/information about the sequence (similar to the frequent words like `a', `the' in English sentences). On the other hand, these common \sqGs\ will contribute to many false positives and consequently increase the running time of subsequent steps.  It is inevitable that in this pruning procedure some true positives are also filtered out.  However, we have observed that by appropriately choosing the filtering threshold $\eta$,\footnote{In our experiments, we choose $\eta = 3 \cdot 10^{-5}$ for \ecoli\ dataset, and $\eta = 10^{-4}$ for \human\ and \scere\ datasets.} we can significantly reduce the number of false positives at the cost of introducing a small number of false negatives.

After the filtering step we perform a subsampling of an $\alpha$-fraction of \qGs\ using the random hash function $\Pi$ (Line~\ref{ln:a-31}). We then only focus on these sampled \qGs\ when measuring the string similarity.  The purpose of performing such a subsampling is to reduce the total running time of the verification step (Line~\ref{ln:a-51}) by producing a set of smaller matching lists $\M_{i,j}$ (Line~\ref{ln:a-32}). On the other hand, it will not affect the accuracy of the algorithm by much.  This is because in the verification step we will consider a pair of input strings $(x_i, x_j)$ who have at least $C$ matched \qG\ pairs, and subsampling \qGs\ by a ratio of $\alpha$ corresponds to subsampling the matched \qG\ pairs by a ratio of $\alpha^2$. Therefore we can scale the threshold $C$ correspondingly to obtain a similar set of candidate string pairs.

\begin{algorithm}[t]
\caption{Search-Similar-$q$-Grams($\delta$, $D$)}
\label{alg:search}
\begin{algorithmic}[1]
\Require $\delta = (s, t, r, i, p)$: a signature for \qG\ $s$ (see Definition~\ref{def:signature} for detailed explanation of the parameters); 

$D$: a table with buckets indexed by $t$; 

\Ensure  $\L \gets \{ (i', p, p')\ |\ \exists \delta' = (s', t, r, i', p') \in D \ s.t.\ \ED(s, s') \le K\}$ 
\State  $\L \leftarrow \emptyset $   
\ForEach{ $\delta' = (s', t, r, i', p')$ stored in $D(t)$}  \label{ln:b-1}
	\If{$\ED(s, s') \le K$} \label{ln:b-2}
	\State $\L \leftarrow  \L \cup (i', p, p')$
	\EndIf
\EndFor
\State Add $\delta$ to the $D(t)$   \label{ln:b-3}
\State \Return $\L$
\end{algorithmic}
\end{algorithm}

We next try to find for each pair of input strings $(x_i, x_j)$, their set of matching \qGs\ (Line~\ref{ln:a-3}-\ref{ln:a-4}). This is done by calling a subroutine Algorithm~\ref{alg:search} to find for each \qG, a list of its matching \qGs\ (with edit distances less than or equal to $K$). More precisely, in Algorithm~\ref{alg:search} we try to find for a \qG\ $s$ an (incomplete) list of matching \qGs\ $s'$ by considering all \qG\ $s'$ such that the corresponding \sqGs\ of $s$ and $s'$ fall into the same bucket in table $D$ (Line~\ref{ln:b-1}). We then perform a brute-force edit distance computation (Line~\ref{ln:b-2}) to make sure that $\ED(s, s')\le K$; if this holds then we record the pair and the positions of the match into $\L$. Finally at Line~\ref{ln:b-3} we add the signature of $s$ into table $D$ to build the table $D$ gradually while performing the search.

\subsection{Verification}
\label{sec:verification}

In this section we discuss how to verify whether a pair of input strings $(x, y)$ overlap at a significant length given a list of their matching \qGs, and if it is the case, what are the shared substrings in the respective strings. For this purpose we employ two subroutines: Algorithm~\ref{alg:verification} performs a basic verification, and outputs a pair of positions on $x$ and $y$ inside the shared substrings if $(x, y)$ is considered as an overlapping pair. We then use Algorithm~\ref{alg:find-location} to recover the actual shared substrings.

\begin{algorithm}[t]
\caption{Verify($x_i$, $x_j$, $\M$)}
\label{alg:verification}
\begin{algorithmic}[1]
\Require $x_i, x_j$: two input strings;  

$\M =\{(u, v)\}$: set of pairs of matched \qG\ positions in $x_i$ and $x_j$;  


\Ensure $o$: reference offset

$pos$: reference position

\State $I \leftarrow \emptyset$ \label{ln:c-1}
\ForEach{$(u, v) \in \M$}
	\State $I \leftarrow I \cup [u - v - \frac{\epsilon}{2} \cdot L, u - v  + \frac{\epsilon}{2} \cdot L]$
\EndFor \label{ln:c-2}
\State Find a value $o$ s.t. $\abs{\{[a,b]\ |\ o \in [a, b], [a, b] \in I\}}$ is maximized \label{ln:c-3}
\State Remove all pairs $(u, v) \in \M$ s.t. $u - v < o -  \frac{\epsilon}{2} \cdot L$ or $u - v > o + \frac{\epsilon}{2} \cdot L$  \label{ln:c-4}

\State $J \leftarrow \emptyset$ \label{ln:c-5}
\ForEach{$(u, v) \in \M$}
	\State $J \leftarrow J \cup [u - \frac{L}{2}, u + \frac{L}{2}]$
\EndFor
\State Find a value $pos$ s.t. $\abs{\{[a,b]\ |\ pos \in [a, b], [a, b] \in J\}}$ is maximized \label{ln:c-6}
\State Remove all pairs $(u, v) \in \M$ s.t. $u  < pos - \frac{L}{2}$ or $u  > pos + \frac{L}{2}$ \label{ln:c-7}

\If{$\abs{\M} < C$} \label{ln:c-8}
	\State \Return $null$
\Else
	\State \Return $(o, pos)$
\EndIf \label{ln:c-9}
\end{algorithmic}
\end{algorithm}

\begin{algorithm}[t]
\caption{Find-Shared-Substrings($x_i$, $x_j$, $o$, $pos$, $\Delta_i$, $\Delta_j$)}
\label{alg:find-location}
\begin{algorithmic}[1]
\Require $x_i, x_j$: two input strings; 

$o$: reference offset; 

$pos$: reference position;  

$\Delta_i, \Delta_j$: sets of \qG\ signatures of $x_i$ and $x_j$

\Ensure $(p_i^s, p_i^e, p_j^s, p_j^e)$: $x_i[p_i^s, p_i^e]$ and $x_j[p_j^s, p_j^e]$ are shared substrings in $x_i$ and $x_j$

\State $\M \leftarrow \{(p, p')\ |\ (s, t, r, i, p) \in \Delta_i,  (s', t, r, j, p') \in \Delta_j, ED(s, s') \le K\}$ 

\State $\Q \leftarrow \{(p, p') \in \M \ |\  p \in [pos - \frac{L}{2}, pos  + \frac{L}{2}], (p - p') \in [o - \frac{\epsilon}{2} \cdot L, o + \frac{\epsilon}{2} \cdot L] \}$ \label{ln:d-2}

\State $(p_i^s, p_j^s) = \arg\min_{(p, p') \in Q} p$, $(p_i^e, p_j^e) =\arg\max_{(p, p') \in Q} p$

\State Remove $(p, p') \in \M$ s.t. $p \in [pos - \frac{L}{2}, pos  + \frac{L}{2}]$ from $\M$  \label{ln:d-3}

\State Sort matches $(p, p') \in \M$ using $\max(p - p_i^e, p_i^s - p)$ in the increasing order \label{ln:d-4}

\ForEach { $(p, p') \in \M$} \label{ln:d-5}
	\If{$0 < p - p_i^e < L \wedge |(p - p') - (p_i^e - p_j^e)| < \epsilon \cdot (p - p_i^e) $  }
		\State $(p_i^e, p_j^e)  \leftarrow (p, \max(p', p_j^e))$
	\EndIf
	\If{$0 < p_i^s - p < L \wedge |(p - p') - (p_i^s - p_j^s)| < \epsilon \cdot (p_i^s - p)$}
		\State $(p_i^s, p_j^s)  \leftarrow (p, \min(p', p_j^s))$
	\EndIf
\EndFor \label{ln:d-6}
\end{algorithmic}
\end{algorithm}

We now describe Algorithm~\ref{alg:verification} and  Algorithm~\ref{alg:find-location} in words.   Let $\M$ be the list of starting positions of the matching pairs of \qGs\ of input strings $x_i$ and $x_j$. We construct bipartite graph $G_{i,j}$ with characters of $x_i$ as nodes on the left side, and characters of $x_j$ as nodes on the right side.  For each matching pair $(u, v)$, there is an edge connecting $x_i[u]$ and $x_j[v]$. For convenience, we slightly abuse the notation by using $(u, v)$ to denote the edge between $x_i[u]$ and $x_j[v]$, and call $(u - v)$ the {\em shift} of the edge.  

It is not hard to imagine that if $x_i$ and $x_j$ overlap, there must be a large cluster of edges of similar shifts in $G_{i,j}$.  Algorithm~\ref{alg:verification} consists of two filtering steps.
In the first step we try to identify a good reference shift $o$ (Line~\ref{ln:c-1}-\ref{ln:c-3}), and remove all the edges whose shifts are far away from $o$ (Line~\ref{ln:c-4}) (more precisely, those pairs $(u, v)$ with $\abs{(u - v) - o} > \frac{\eps}{2} \cdot L$).  According to the previous literature,  SMRT sequencing reads have accuracy $82\%-88\%$ \cite{K15}. We thus set the error tolerance rate $\epsilon$ to be $0.2$. 

After finding a good reference shift, we try to find a dense area (or simply, a reference position $pos$ in $x_i$) which contains many edges whose shifts are close to $o$ (Line~\ref{ln:c-5}-\ref{ln:c-6}).  We then remove all the edges that are not in this dense area (Line~\ref{ln:c-7}). Finally, we count the number of edges in the dense areas; if the number is at least $C$, then we consider $(x_i, x_j)$ an overlapping pair and return the reference edge (determined by $o$ and $pos$); otherwise we simply return $null$ (Line~\ref{ln:c-8}-\ref{ln:c-9}).

We should note that all of these operations are performed on a subset $\M$ of matched \qG\ pairs in $x_i$ and $x_j$.  By ``subset'' we mean that $\M_{i,j}$ is constructed after the subsampling step at Line~\ref{ln:a-31} in Algorithm~\ref{alg:overlap}. As mentioned above, the purpose of the subsampling is to reduce the running time in the verification step.  In contrast, when the actual shared substrings between $x_i$ and $x_j$ are found by Algorithm~\ref{alg:find-location}, we exploit the {\em complete} set of matched \qG\ pairs, which will not significantly increase the overall running time because after verification, the number of input string pairs becomes much smaller.

Now, we turn to the details of the algorithm for determining the actual shared substrings between $x_i$ and $x_j$ (Algorithm~\ref{alg:find-location}).  We again first construct the list $\M$ of matching \qGs.  This can be done by a synchronized linear scan on the two sets $\Delta_i$ and $\Delta_j$, after sorting the tuples by their $r$ values. Next, starting from the reference edge determined by $o$ and $pos$, we first locate the corresponding dense areas (Line~\ref{ln:d-2}-\ref{ln:d-3}). We then try to extend this dense area by adding one by one the matching edges outside this dense areas but still within a distance of $L$ from the dense area, in the increasing order of the distances between these matching edges to the dense area (Line~\ref{ln:d-4}-\ref{ln:d-6}).  Finally the algorithm returns the extended area as the shared substrings between $x_i$ and $x_j$.




\section{Experiments}
\label{sec:exp}

In this section we present experimental studies of \sqG\ and its application to detect overlaps among SMRT sequencing reads.

\subsection{Tested Algorithms} 
We have implemented our algorithms presented in previous sections in C++, and complied them using GCC 5.4.0 with O3 flag.

To facilitate the investigation of properties of \sqGs, we introduce an additional algorithm named {\em Find-Similar-$q$-Gram-Pairs}, which uses the \sqG\ technique to find pairs of input \qGs\ whose edit distances are at most $K$ for a given distance threshold $K$.  The algorithm is depicted in Algorithm~\ref{alg:join}.  Let us describe it in words briefly.  Essentially, {\em Find-Similar-$q$-Gram-Pairs} can be seen as running {\em Search-Similar-$q$-Grams} (Algorithm~\ref{alg:search}) for each input \qG.  Of course in this investigation we do not need to carry the data structure $\delta(s, \cdot, \cdot, \cdot, \cdot)$ for each \qG\ $s$ that we used  in Algorithm~\ref{alg:search} (for the application of overlap detection).  Moreover, as mentioned at the end of Section~\ref{sec:q-gram}, we can choose to repeat the CGK-embedding and the subsampling for $d$ and $z$ times respectively, so that for each \qG\ $s$ we create $d \cdot z$ \sqGs. By doing this we can generate more similar \qG\ pairs which can be used to potentially boost the accuracy of our application. 
We will test Algorithm~\ref{alg:join} for various $d$ and $z$ values. While in our applications in Section~\ref{sec:exp-compare} we only perform the embedding and the subsampling {\em once}, which is enough for obtaining good accuracy.

\begin{algorithm}[t]
\caption{Find-Similar-$q$-Gram-Pairs($\S$, $d$, $z$)}
\label{alg:join}
\begin{algorithmic}[1]
\Require $\mathcal{S} = \{s_1, \ldots, s_n\}$:  set of $q$-grams;

$d$: number of CGK-embeddings; 

$z$: number of subsamplings;  

\Ensure  $\O \gets \{(s_i, s_j)\ |\ s_i, s_j \in \S, i \neq j, \text{ED}(s_i, s_j) \le K\}$ 

\State $\mathcal{C}  \leftarrow \emptyset$ 

\ForEach{$j \in [d]$}
	\State Pick a random string $R_c^j$ from $\{0,1\}^{\kappa \abs{\Sigma}}$,
	\ForEach{$k \in [z]$}
		\State Pick a random string $R_s^k$ from $\{0,1\}^\kappa$ under the constraint that it contains $m$ $1$-bit
		\State Initialize a new table $D^{jk}$
		\ForEach{$i \in [n]$}
			\State $t_i^{jk} \leftarrow$ Generate-Smooth-$q$-Gram ($s_i$, $R_c^j$, $R_s^k$)
		\EndFor
		\State Count for each distinct \sqG\ its frequency
		\ForEach{$i \in [n]$}
			\If{frequency of $t_i^{jk}$ is less than $\eta \cdot n$}
			\ForEach{$q$-gram $s$ stored in the $D^{jk}(t_i^{jk}) $}
				\State $\mathcal{C}  \leftarrow \mathcal{C}  \cup (s, s_i)$ 
			\EndFor
			\State Store $s_i$ in $D^{jk}(t_i^{jk})$
			\EndIf
		\EndFor
	\EndFor
\EndFor
\State Remove duplicate pairs in $\mathcal{C}$ 	
\ForEach {$(x, y) \in \mathcal{C}$}
	\If{$\text{ED}(x, y) \le K$}
	\State $\mathcal{O} \leftarrow  \mathcal{O} \cup (x, y)$
	\EndIf
\EndFor
\end{algorithmic}
\end{algorithm}

In Section~\ref{sec:exp-compare} we compared Algorithm~\ref{alg:overlap} with existing overlap detection algorithms.  For convenience, we call our algorithm \our. We briefly describe each of the competitors below.  

\mhap\cite{BKC15}\footnote{Implementation obtained from \url{https://github.com/marbl/MHAP}}: this algorithm generates \qGs\ of all sequences and then filters out those with frequencies greater than $0.00001$ times the total number of \qGs. Next, it uses multiple {\em Minhash} \cite{B98} functions to find matching \qGs\ between sequences, and then select pairs of sequences that have at least $3$ matching \qGs\ as candidate pairs.  For each candidate pair, it uses a modified sort-merge algorithm to find more accurate \qG\ matches, and then computes the boundary of the overlap region using a uniformly minimum-variance unbiased (UMVU) estimator \cite{CA83}.  

\mini \cite{L16}\footnote{Implementation obtained from \url{https://github.com/lh3/minimap}}: this algorithm generates \qGs\ of all sequences and then filters out the top $0.001$ fraction of the most frequent ones.  Next, it hashes each \qG\ to a value in $\Sigma^q$, and selects \qGs\ with the smallest hash values in every $5$ consecutive \qGs\ as signatures of the input sequence.  It then find all matching signatures between input sequences; pairs of sequences that have at least one shared signature are identified as candidate pairs.  \mini\ then calculates a cluster of \qG\ matches for each candidate pair, and then finds a maximum colinear subset of matches by solving a longest increasing sequence problem.  If the size of the subset is larger than $4$, then \mini\ computes and outputs the overlap region using the subset of matches.

\dalign \cite{CT12}\footnote{Implementation obtained from \url{https://github.com/thegenemyers/DALIGNER}}:  this algorithm generates \qGs\ of all sequences and then filters out those that occur more than $100$ times.  It then considers all the remaining $q$-grams directly and computes all the matching \qGs\ between pairs of sequences.  Pairs of sequences with at least one shared \qG\ are identified as a candidate. Next, for each candidate, it uses a linear time difference algorithm \cite{M86} to compute a local alignment between the two sequences, and outputs the pair if the alignment length is greater than the given threshold.  

We note that all these algorithms are under the same ``seed-extension'' framework as ours: they first find all the matched \qGs\ between input sequence pairs, and then extend seeds to potential overlaps.  The major difference between our \our\ and the existing tools is that we have {\em relaxed} the strict \qG\ matches to approximate \qG\ matches (via \sqG) to improve the accuracy of the output. Our specifics for overlap detection and determination of shared  substrings are also different from the existing algorithms.

In our experiments we run \our\ with parameters $q = 14, m = 21, \alpha = 0.15, K = 2, C = 3, L = 500, \epsilon = 0.2$, and $\kappa = 2q$
\ifdefined \submission
\else
(except for Figure~\ref{fig:numm}, where we have tested different $\kappa$ values)
\fi
. 
We choose $\eta = 0.00003$ for \ecoli, and $\eta = 0.0001$ for \human\ and \scere.  We run \mhap\ with parameters ``-num-hashes 1256'',  \mini\ with parameters ``-k 15 -Sw5 -L100 -m0 -t8'', and \dalign\ with parameter ``-H500''. All other parameters were selected as default settings.  


We note that the performance of \mini\ is sensitive to the filter threshold (``-f'') it uses. Thus, besides the default parameter, we also choose an alternative parameter ``$-f\ 0.00000001$'' (according to \cite{KB17}'s recommendation), which essentially means that almost no $q$-gram will be filtered out.  Intuitively, such a change will lead to better recall values (but possibly worse precision values) at the cost of greater memory usage and running time.  We call the original version \miniD\ and the new version \miniA.  


\subsection{The Setup}

\paragraph{Datasets} 
We test algorithms using real world datasets from \pac\ SMRT sequencing.\footnote{The data was downloaded from \mhap's supporting data website: \url{http://www.cbcb.umd.edu/software/PBcR/mhap/index.html}} The statistics of these datasets are described in Table~\ref{tab:stat}.  In Section~\ref{sec:exp-join} we test Algorithm~\ref{alg:join} using the dateset \ecolis;  the number of q-grams for ecoli-small is $100 \cdot 4781$. In Section~\ref{sec:exp-compare} we compare different algorithms using much larger datasets \ecoli, \scere\ and \human. 

\begin{table}[t]
\centering 
\begin{tabular}{lcc} 
\hline
Datasets & number of strings &Average Length  \\  \hline 
\ecolis &100  &4781 \\  \hline
\ecoli &46960  &4221 \\  \hline
\scere &48279   &3032  \\  \hline
\human &47535   &3100  \\  \hline
\end{tabular}
\caption{Statistics of Tested Datasets}
\label{tab:stat}
\end{table}

\paragraph{Measurements}
In Section~\ref{sec:exp-join} we report the number of matching \qG\ pairs detected by the Algorithm~\ref{alg:join}. Each result is an average of $5$ runs.  In Section~\ref{sec:exp-compare},  we report four types of measurements in our experiments: {\em recall}, {\em precision}, {\em memory usage} and {\em running time}. We choose the evaluation program used by \mhap\ to calculate precision and recall; all parameters are selected as default except that the evaluation overlap length threshold $\Gamma$ is set to be $500$ or $2000$.  The evaluation program learns the ground truths from the reads to references mappings obtained by \blasr~\cite{CT12}. We note that the evaluation algorithm does not simply use edit distance as the criteria to compute precise and recall. Instead, it maps all the reads to the reference sequences, and computes for each pair of reads their overlap positions and lengths from the mapping results. This evaluation method is widely used in overlap detection because it considers biological meanings of overlaps between reads.

We also present the $F_1$ score: 
$$F_1 = 2\times \frac{\text{precision } \times \text{recall}}{\text{precision } + \text{recall}},$$ 
which is an integrated metric evaluating both precision and recall.

All algorithms use multiple threads in execution; we thus measure the CPU time for comparison.  The memory usage we report is the maximum memory usage of a program during its execution. We note that although  all the tested algorithms are randomized, we use a fixed random seed for all of them to guarantee the consistency among outputs. 

\paragraph{Computing Environment}
All experiments are conducted on a Dell PowerEdge T630 server with 2 Intel Xeon E5-2667 v4 3.2GHz CPU with 8 cores each, and 256GB memory.

\subsection{Finding \qG\ Pairs with Small Edit Distance}
\label{sec:exp-join}

\ifdefined \submission
In this section we present the performance of {\em Find-Similar-$q$-Gram-Pairs}.
We choose the parameters for Algorithm~\ref{alg:join} from Table~\ref{tab:para}; parameters underlined are default values.  

\else
In this section we present the performance of {\em Find-Similar-$q$-Gram-Pairs} (Algorithm~\ref{alg:join}).
We choose the parameters for Algorithm~\ref{alg:join} from Table~\ref{tab:para}; parameters underlined are default values.  
\fi

\begin{table}[]
\centering 
\begin{tabular}{lcc} 
\hline
&Parameter &Values \\  \hline 
&$q$ &$12, 14, 16$\\  \hline
&$K$  &$1, 2$\\  \hline
&$m$ ($\times q$) &$1, 1.25,  \underline{1.5}, 1.75, 2, 3$\\  \hline
&$d$ &$\underline{1}, 2, 3, 4, 5$\\  \hline
&$z$ &$\underline{1}, 2, 3, 4, 5$\\  \hline
&$\eta$ &$5\times10^{-6}, 1\times10^{-5}, 5\times10^{-5}, 1\times10^{-4}, \underline{1}$
\end{tabular}
\caption{Parameters for Algorithm~\ref{alg:join}
}
\label{tab:para}
\end{table}

\paragraph{Matching Pairs of \qGs}  We first study how different parameter values $m$ (the length of the \sqG), $d$ (the number of CGK-embeddings), and $z$ (the number of samplings) influence the number of almost matching \qG\ pairs that we can find.  
\ifdefined \submission
Our results are presented in Figure~\ref{fig:new-num}. Due to the space constraints we only show the results for $q = 14$, and leave those for other $q$ values to the full version of this paper; the results are all similar.

\begin{figure*}[t]
\begin{minipage}[d]{0.33\linewidth}
\centering
\includegraphics[width=0.8\textwidth]{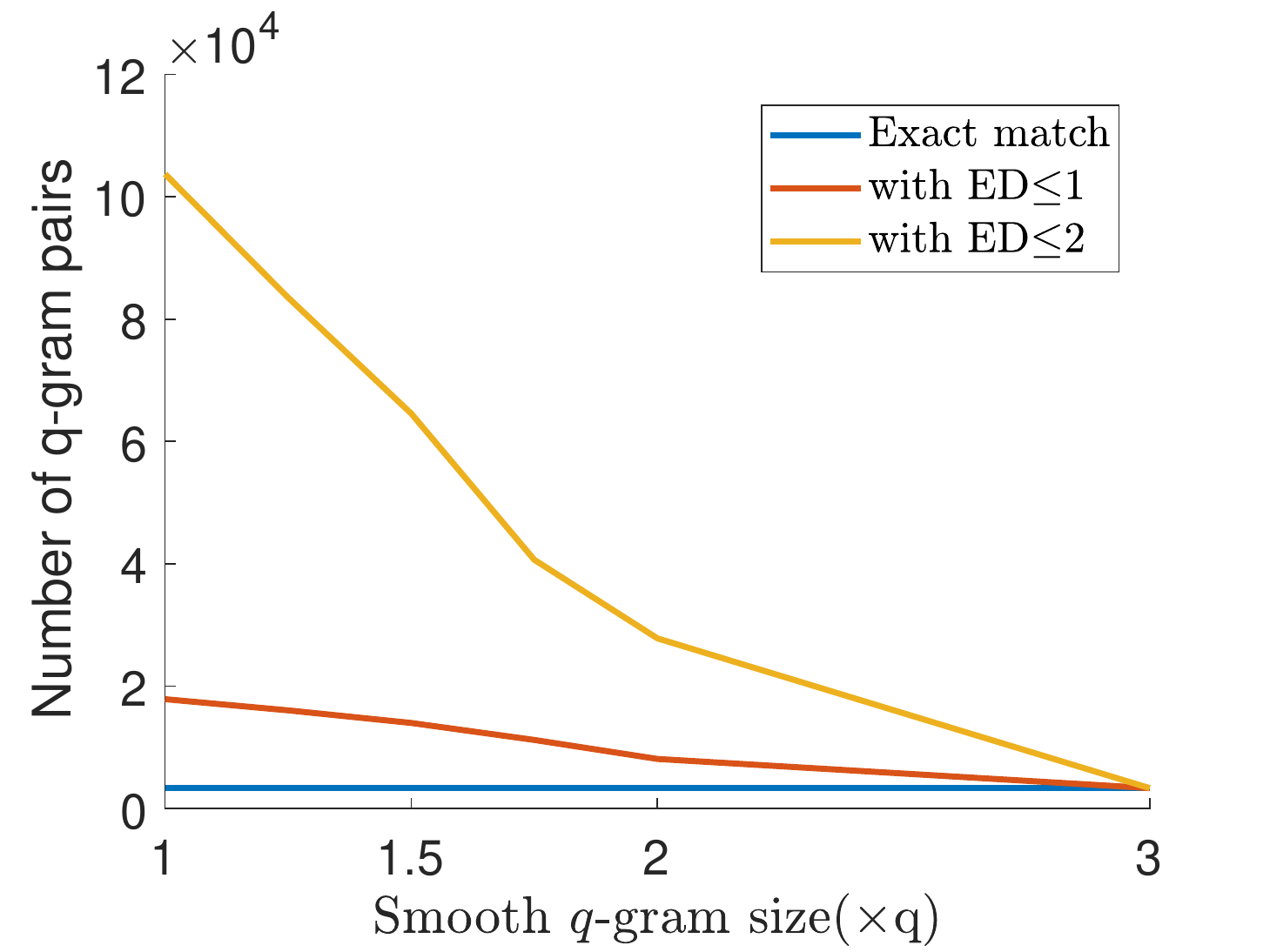}
\end{minipage}
\begin{minipage}[d]{0.33\linewidth}
\centering
\includegraphics[width=0.8\textwidth]{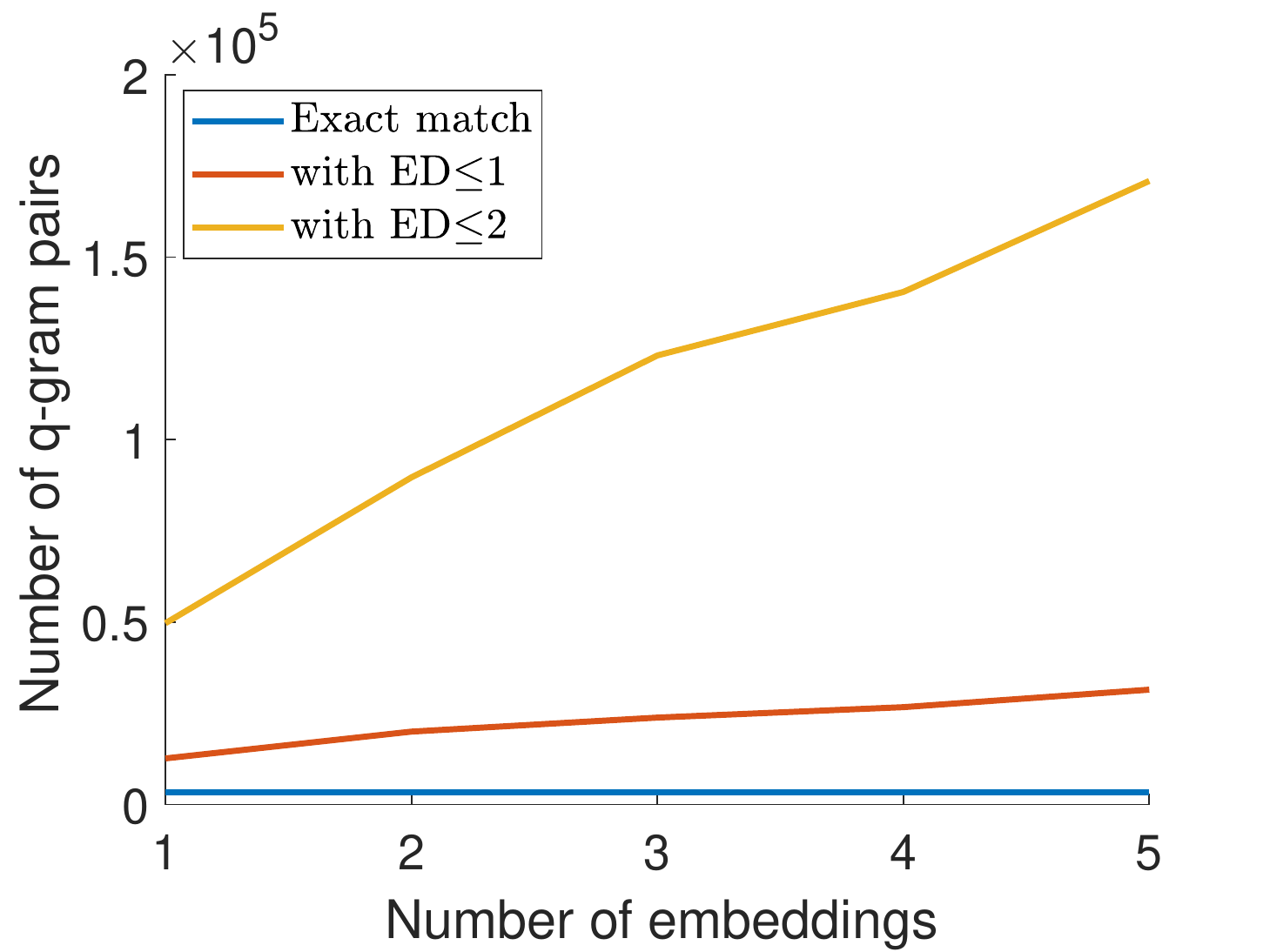}
\end{minipage}
\begin{minipage}[d]{0.33\linewidth}
\centering
\includegraphics[width=0.8\textwidth]{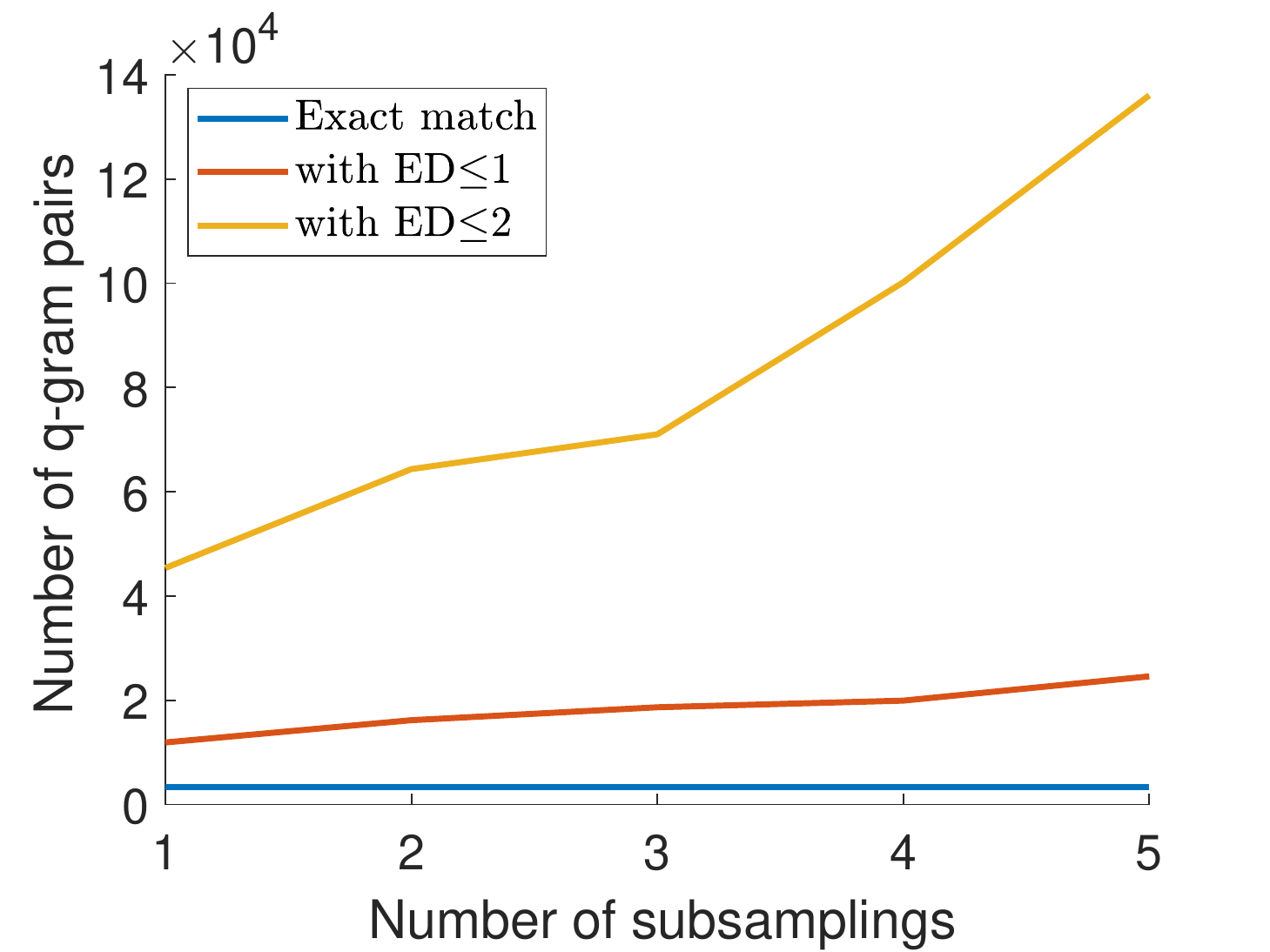}
\end{minipage}
\caption{Number of matching \qG\ pairs vs smooth \qG\ size $m$, number of embeddings $d$, and number of subsamplings $z$;  $q = 14$; on \ecolis}
\label{fig:new-num}
\end{figure*}

\else
Our results are presented in Figures~\ref{fig:numm}, \ref{fig:numr} and \ref{fig:numz}.  


\begin{figure*}[t]
\begin{minipage}[d]{0.33\linewidth}
\centering
\includegraphics[width=0.8\textwidth]{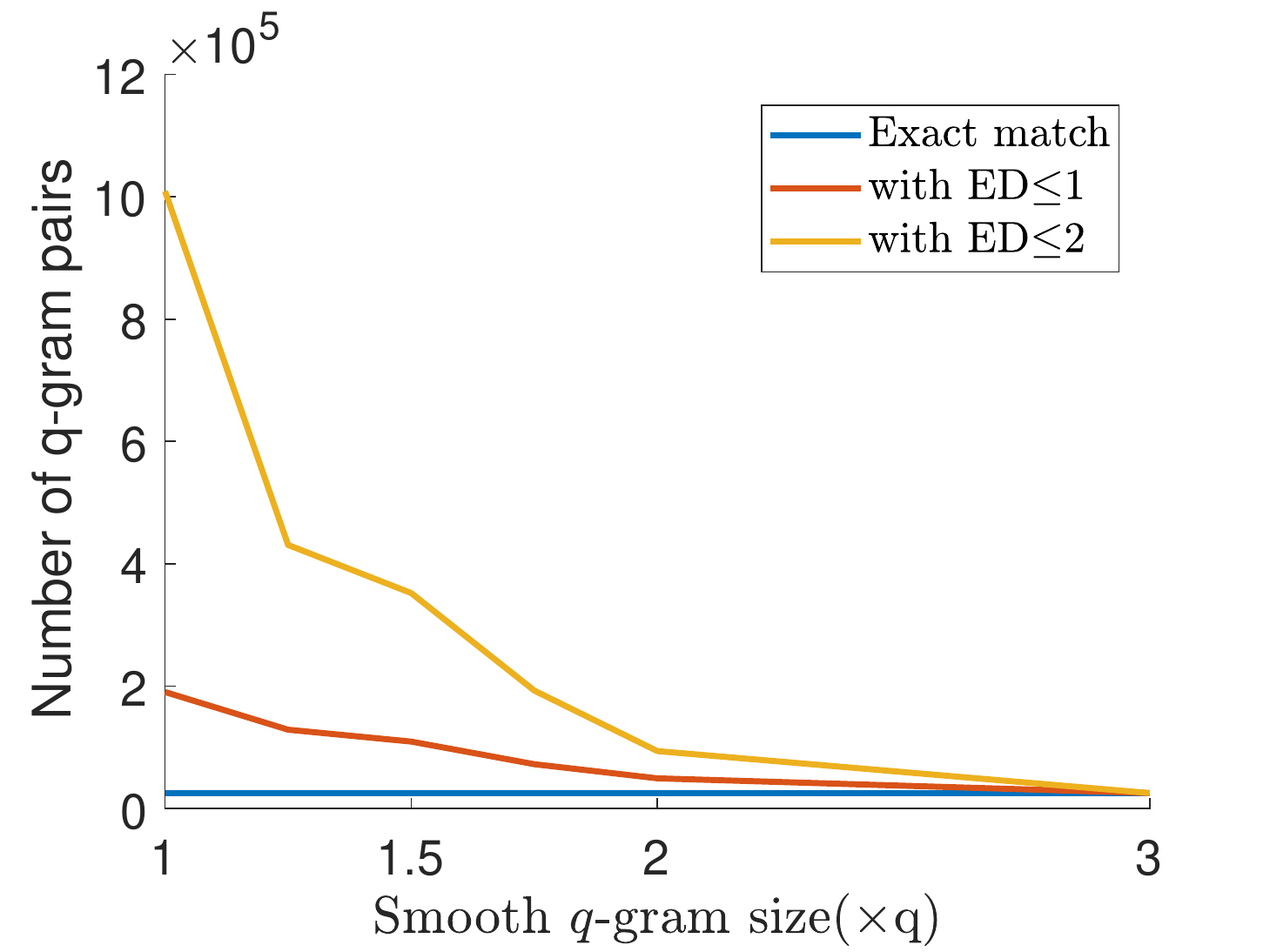}
\centerline{$q = 12$}
\end{minipage}
\begin{minipage}[d]{0.33\linewidth}
\centering
\includegraphics[width=0.8\textwidth]{sr14.pdf}
\centerline{$q = 14$}
\end{minipage}
\begin{minipage}[d]{0.33\linewidth}
\centering
\includegraphics[width=0.8\textwidth]{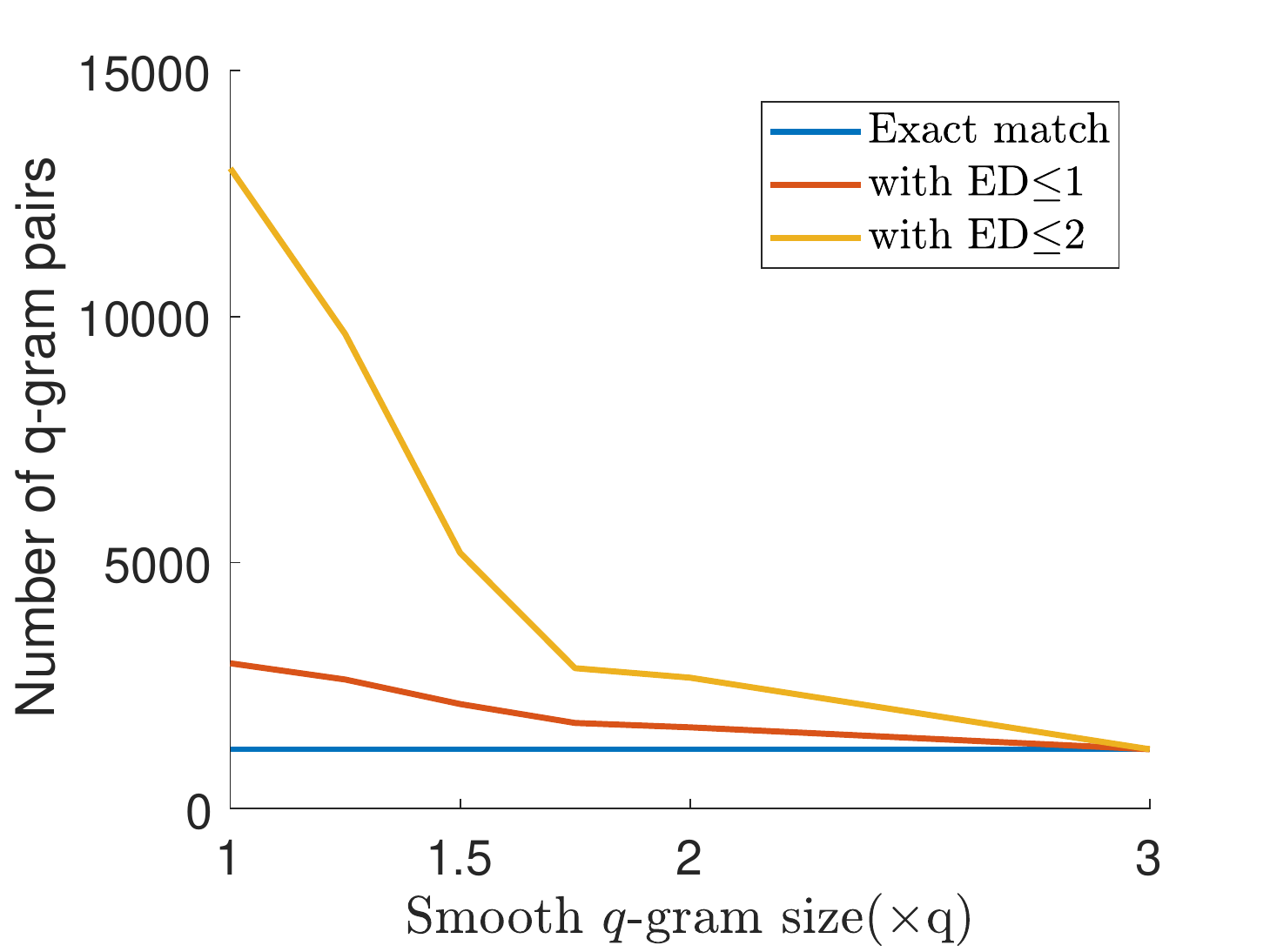}
\centerline{$q = 16$}
\end{minipage}
\caption{Number of matching \qG\ pairs vs smooth \qG\ size $m$; on \ecolis}
\label{fig:numm}
\end{figure*}

\begin{figure*}[t]
\begin{minipage}[d]{0.33\linewidth}
\centering
\includegraphics[width=0.8\textwidth]{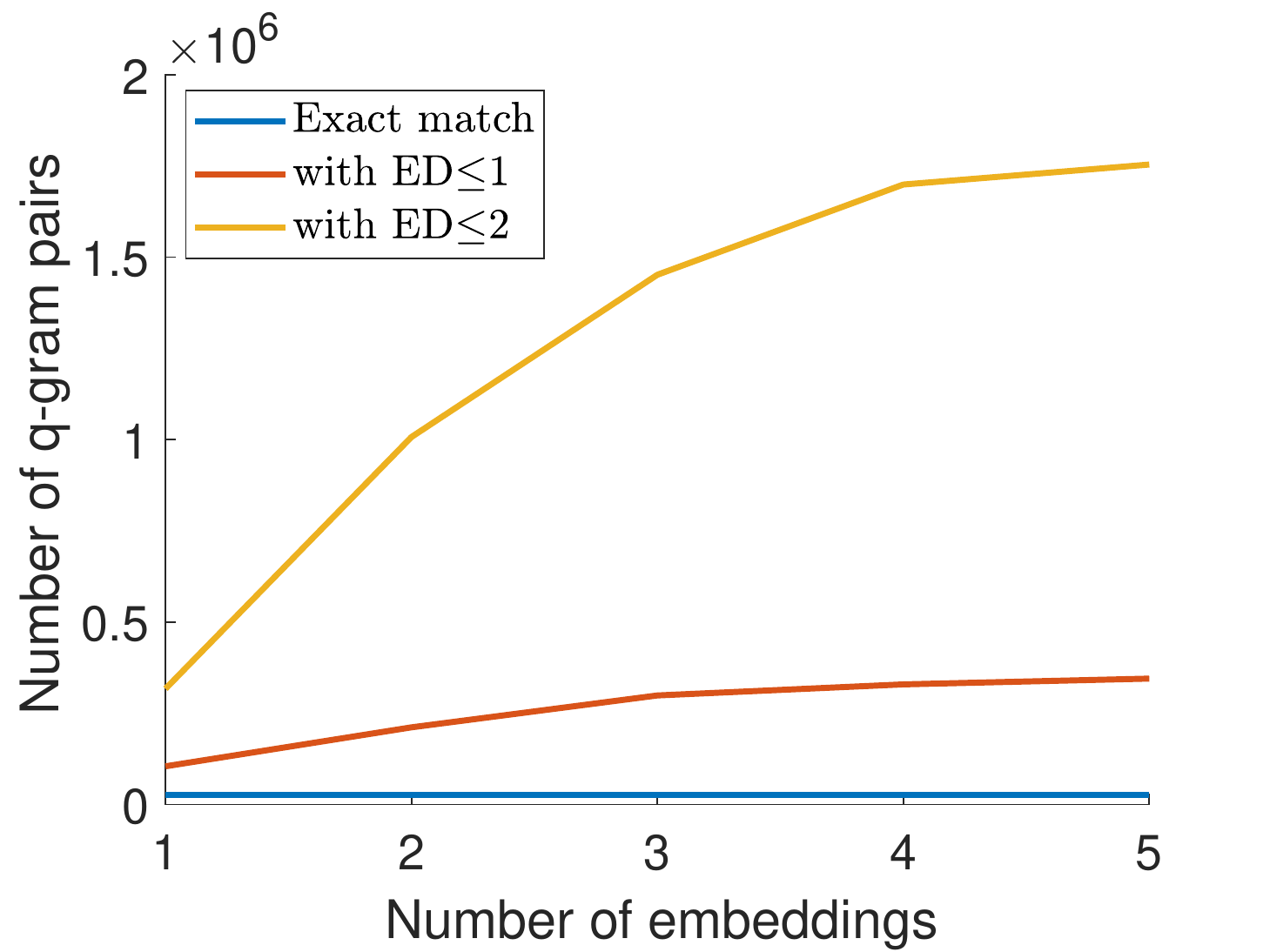}
\centerline{$q = 12$}
\end{minipage}
\begin{minipage}[d]{0.33\linewidth}
\centering
\includegraphics[width=0.8\textwidth]{ebd14.pdf}
\centerline{$q = 14$}
\end{minipage}
\begin{minipage}[d]{0.33\linewidth}
\centering
\includegraphics[width=0.8\textwidth]{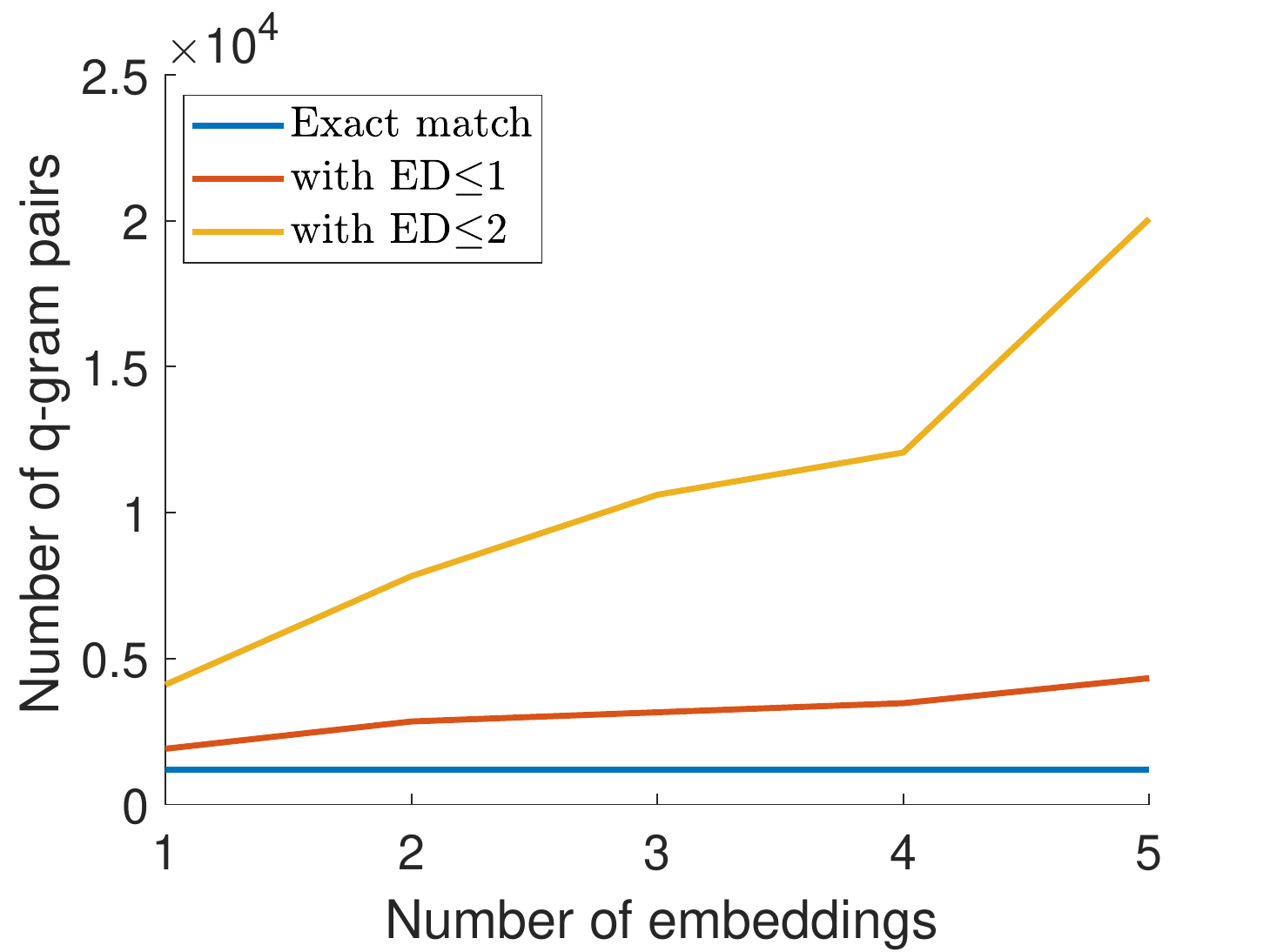}
\centerline{$q = 16$}
\end{minipage}
\caption{Number of matching \qG\ pairs vs number of embeddings $d$; on \ecolis}
\label{fig:numr}
\end{figure*}

\begin{figure*}[t]
\begin{minipage}[d]{0.33\linewidth}
\centering
\includegraphics[width=0.8\textwidth]{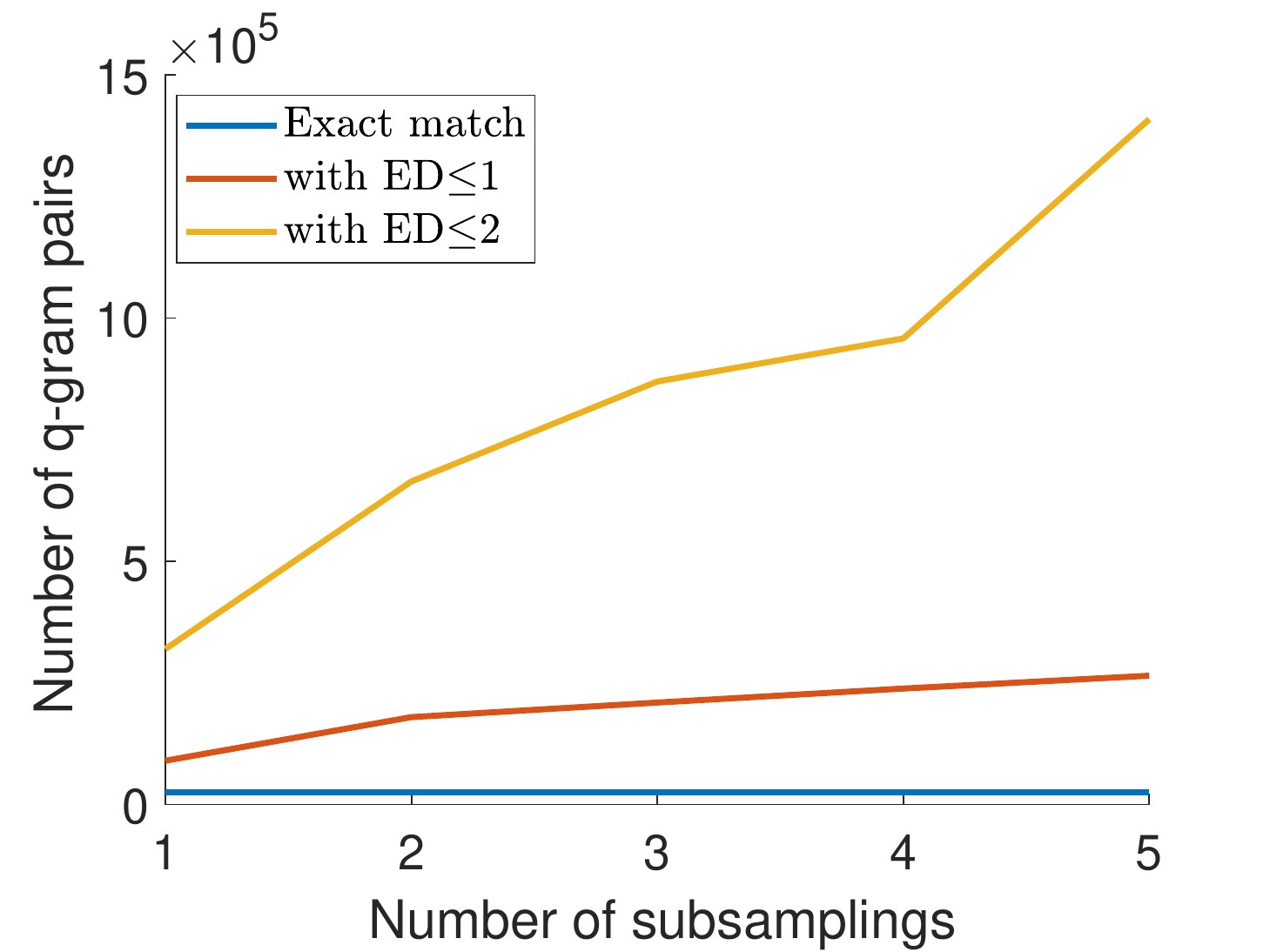}
\centerline{$q = 12$}
\end{minipage}
\begin{minipage}[d]{0.33\linewidth}
\centering
\includegraphics[width=0.8\textwidth]{hash14.pdf}
\centerline{$q = 14$}
\end{minipage}
\begin{minipage}[d]{0.33\linewidth}
\centering
\includegraphics[width=0.8\textwidth]{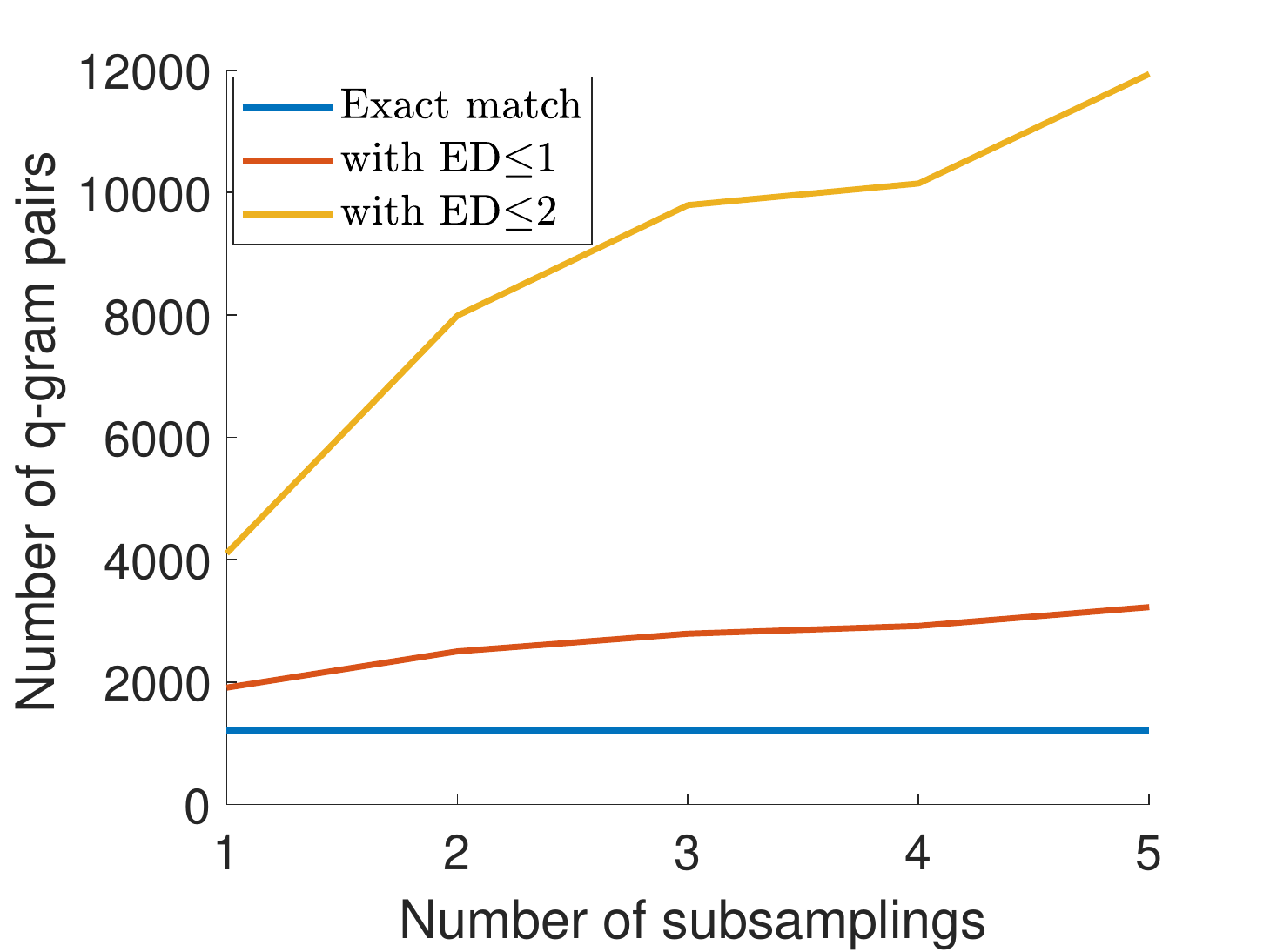}
\centerline{$q = 16$}
\end{minipage}
\caption{Number of matching \qG\ pairs vs number of subsamplings $z$; on \ecolis}
\label{fig:numz}
\end{figure*}
\fi

We observe that the number of matching \qGs\ increases when $m$ decreases, and significantly increases when $d$ and $z$ increase.  For example, fix $q$ to be $14$. For $m = 1.5 \times q$, we can detect $17.4$ times \qG\ matches with edit distance being at most $2$ of that of exact matches, using only one subsampling and one embedding.  With $d = 5$, we could detect $51.3$ times (distinct) \qG\ matches with edit distance being at most $2$ of that of exact matches; and with $z = 5$ subsamplings, we could detect $40.8$ times (distinct) \qG\ matches. 

In the rest of this section we will simply set $d = z = 1$, mainly for the sake of time/space saving.  We found that by setting $d = z = 1$ we can already obtain very good accuracy, though higher $d$ and $z$ values can potentially lead to better accuracy.

\paragraph{True and False Positives}
We next study how different parameters $m, \eta$ influence the number of false positives. We call the pairs of \qGs\ whose edit distances are at most $2$ {\em true positives}, and those with edit distances larger than $2$ {\em false positives}.  
\ifdefined \submission
Our results are presented in Figures~\ref{fig:new-fp}. Due to the space constraints we only show the results for $q = 14$, and leave those for other $q$ values to the full version of this paper; the results are all similar.

\begin{figure*}[t]
\begin{minipage}[d]{0.33\linewidth}
\centering
\includegraphics[width=0.8\textwidth]{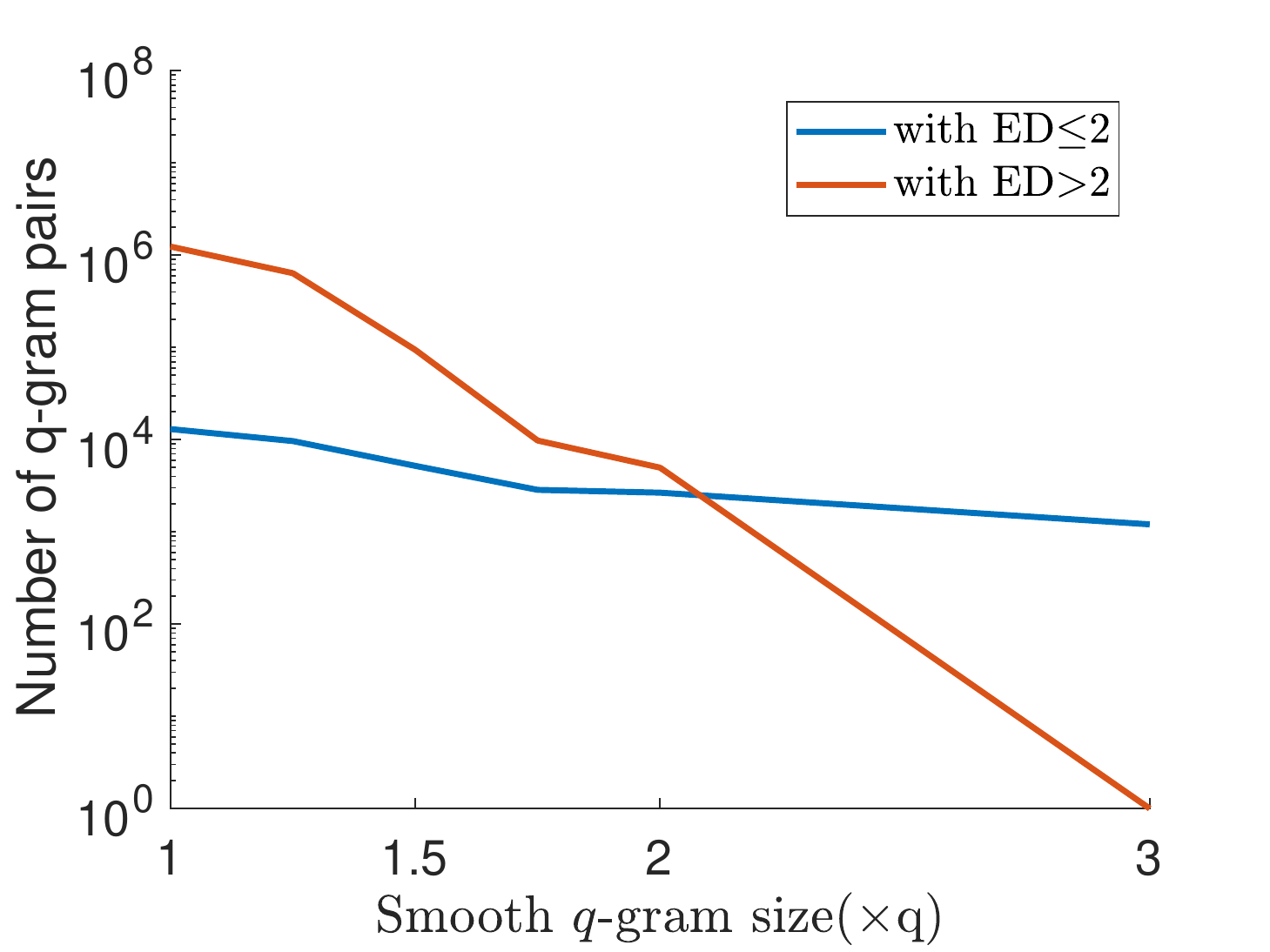}
\end{minipage}
\begin{minipage}[d]{0.33\linewidth}
\centering
\includegraphics[width=0.8\textwidth]{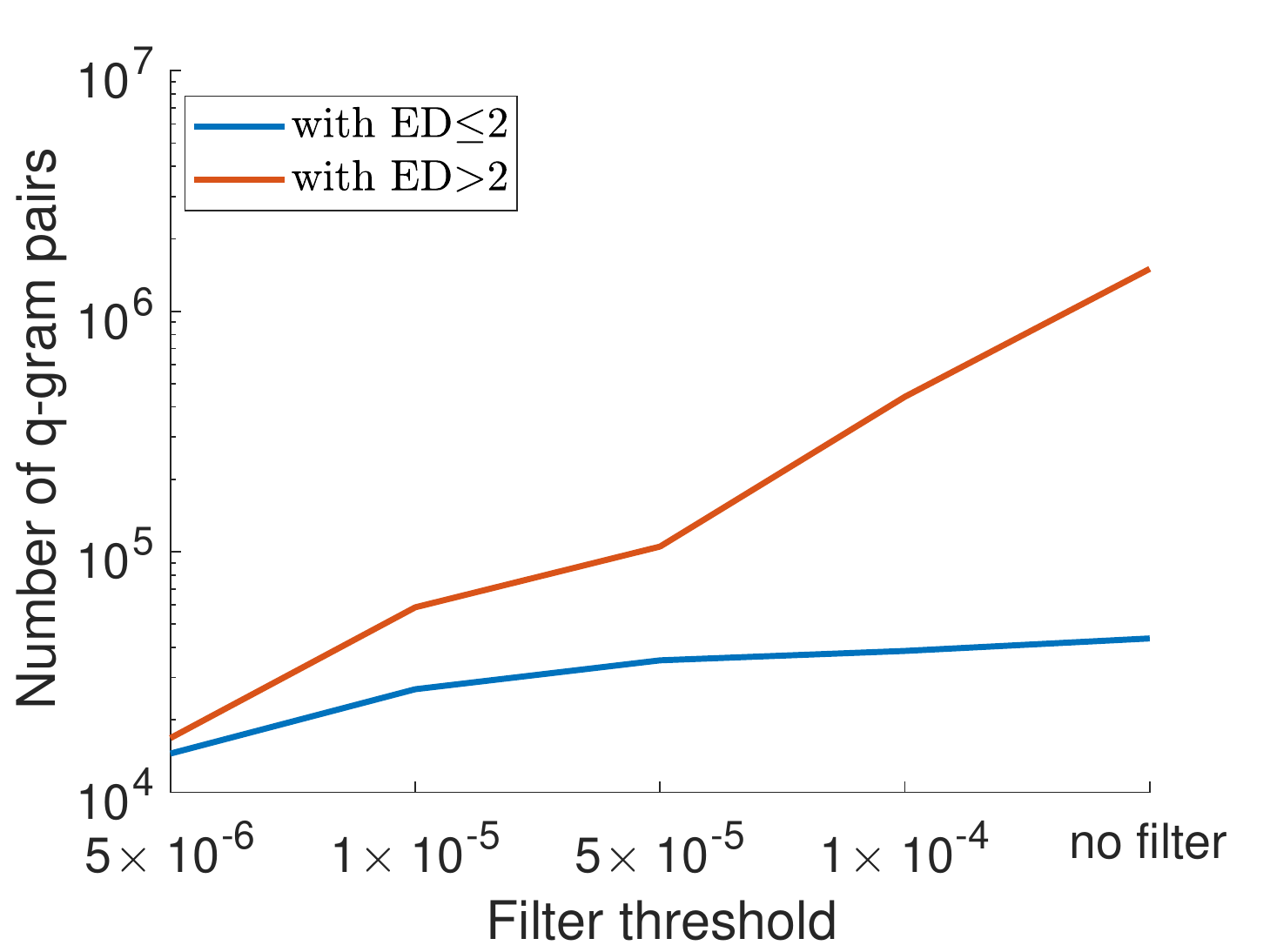}
\end{minipage}
\caption{Number of true/false positive $q$-gram pairs vs smooth $q$-gram size $m$ and filter threshold $\eta$; $q = 14$; on \ecolis}
\label{fig:new-fp}
\end{figure*}

\else
Our results are presented in Figures~\ref{fig:fpm} and \ref{fig:fpT}.  
\begin{figure*}[t]
\begin{minipage}[d]{0.33\linewidth}
\centering
\includegraphics[width=0.8\textwidth]{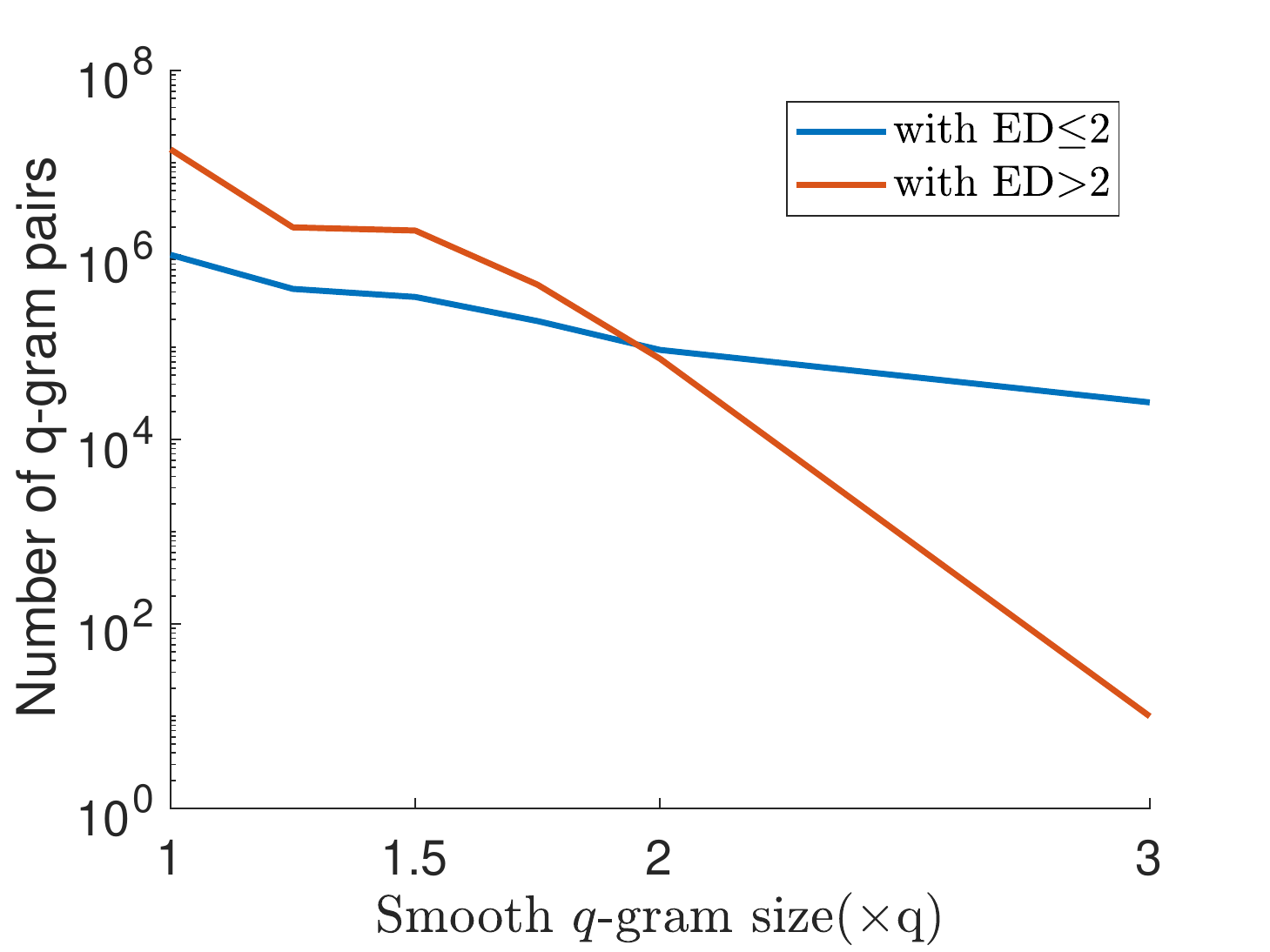}
\centerline{$q = 12$}
\end{minipage}
\begin{minipage}[d]{0.33\linewidth}
\centering
\includegraphics[width=0.8\textwidth]{fp14.pdf}
\centerline{$q = 14$}
\end{minipage}
\begin{minipage}[d]{0.33\linewidth}
\centering
\includegraphics[width=0.8\textwidth]{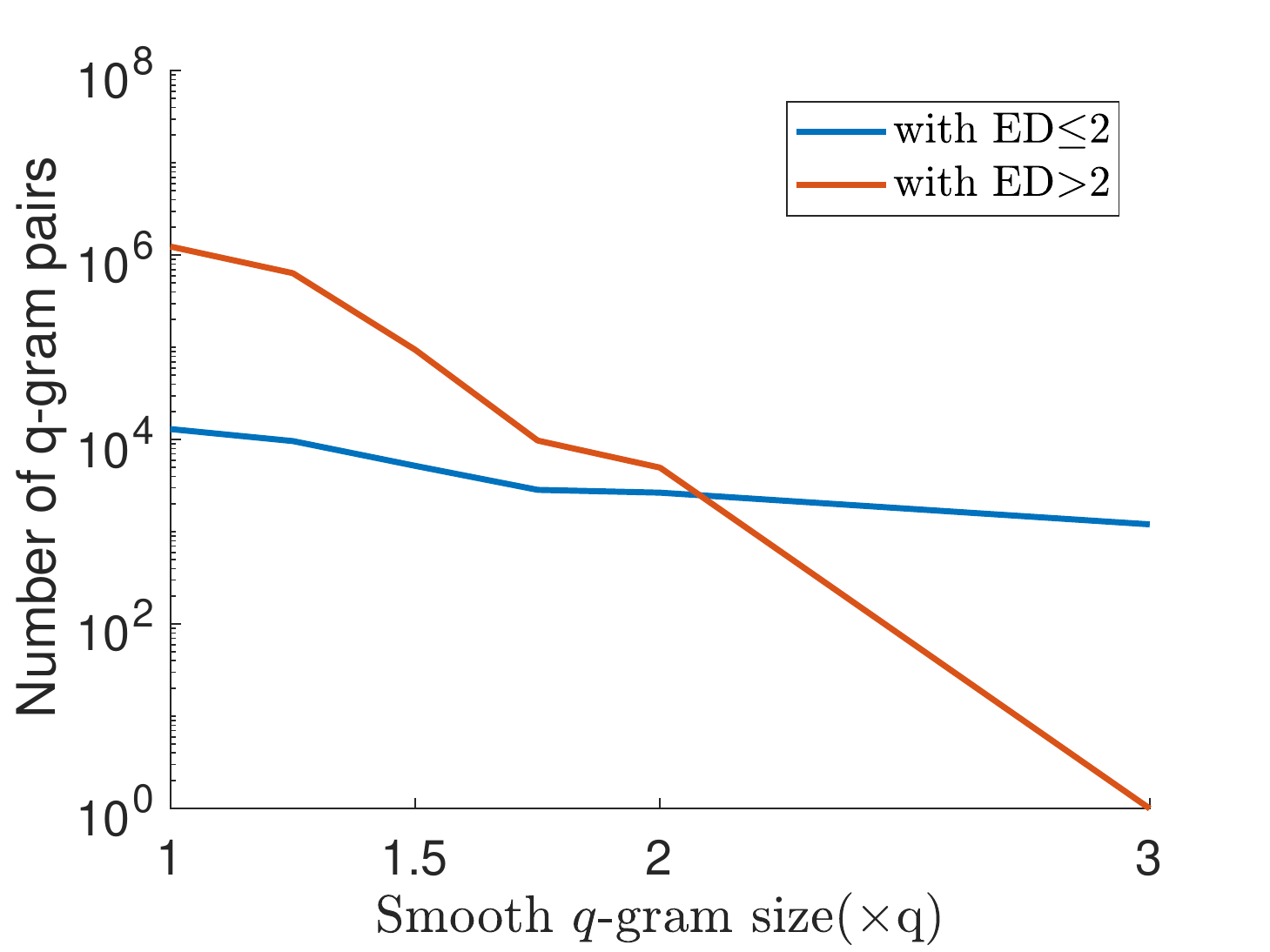}
\centerline{$q = 16$}
\end{minipage}
\caption{Number of true/false positive $q$-gram pairs vs smooth $q$-gram size $m$; on \ecolis}
\label{fig:fpm}
\end{figure*}

\begin{figure*}[t]
\begin{minipage}[d]{0.33\linewidth}
\centering
\includegraphics[width=0.8\textwidth]{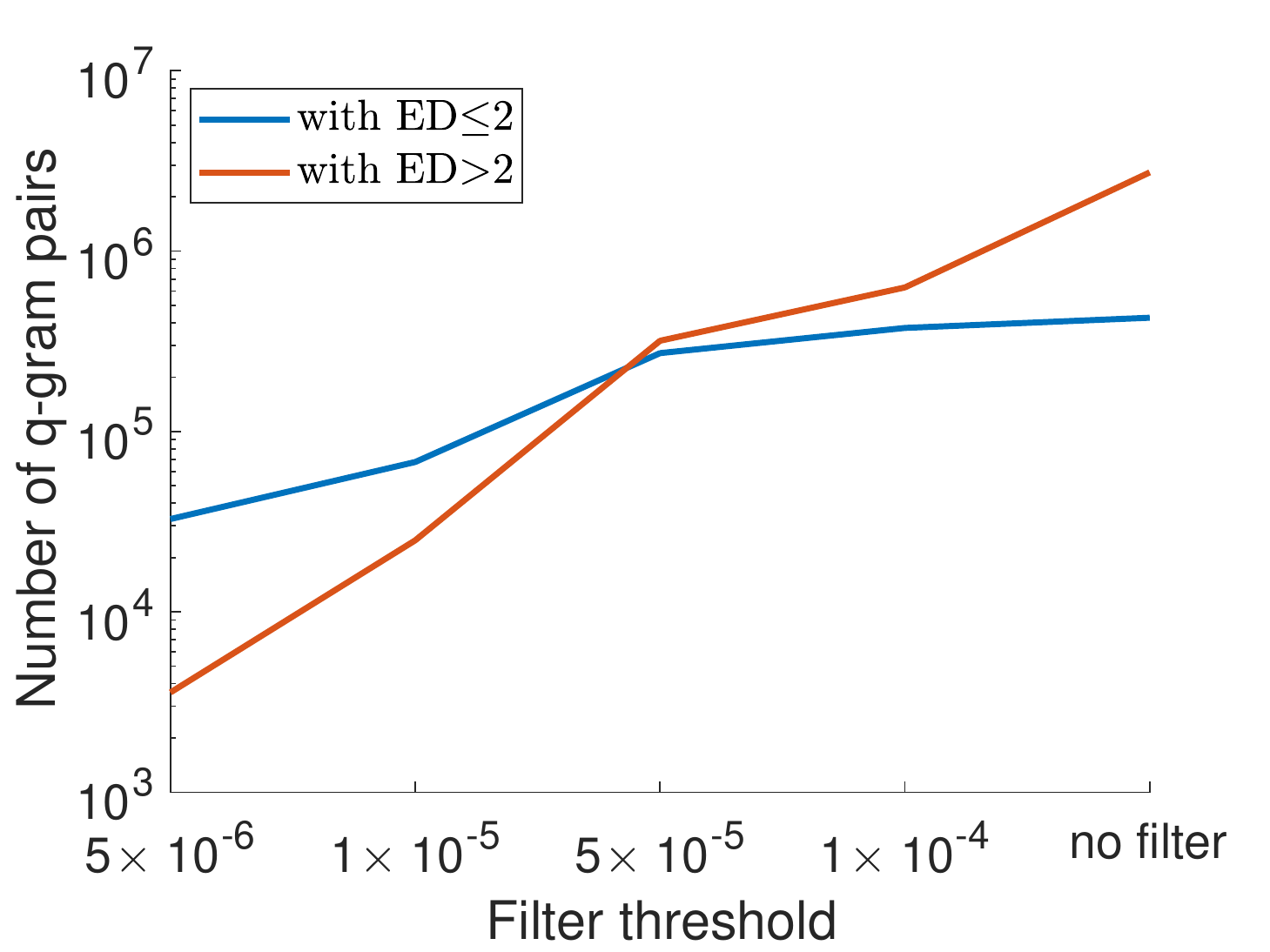}
\centerline{$q = 12$}
\end{minipage}
\begin{minipage}[d]{0.33\linewidth}
\centering
\includegraphics[width=0.8\textwidth]{filter14.pdf}
\centerline{$q = 14$}
\end{minipage}
\begin{minipage}[d]{0.33\linewidth}
\centering
\includegraphics[width=0.8\textwidth]{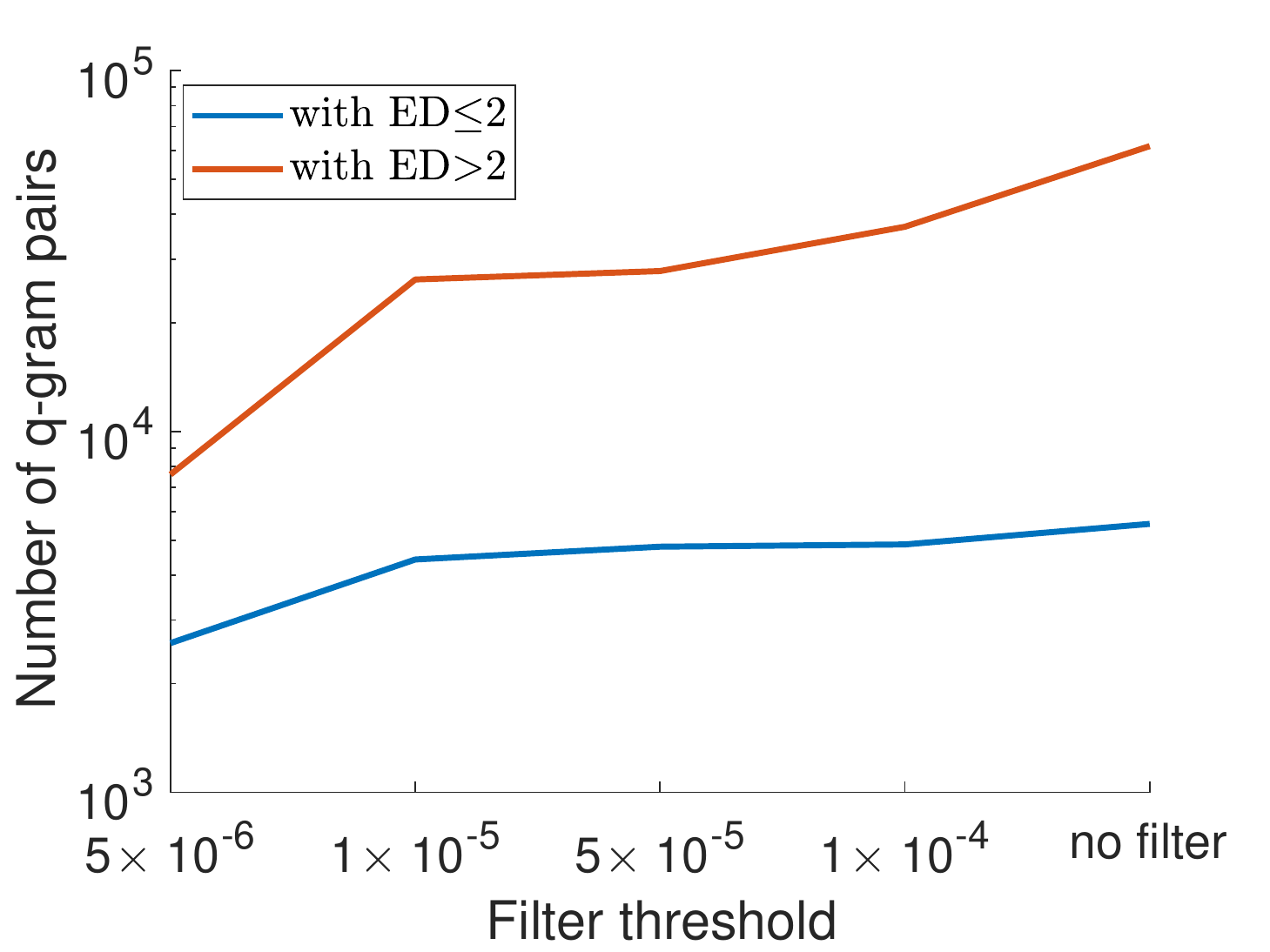}
\centerline{$q = 16$}
\end{minipage}
\caption{Number of true/false positive $q$-gram pairs vs filter threshold $\eta$; on \ecolis}
\label{fig:fpT}
\end{figure*}

We note that Figure \ref{fig:fpm} and Figure \ref{fig:numm} come from the same set of experiments, but with different edit distance ranges and scales recorded.  We observed that the number of false positives increases with $\eta$, and decreases sharply with $m$. 

Figure \ref{fig:numm} and Figure~\ref{fig:fpm} also guide us on how to choose $m$ to balance the number of true positives and false positives.  Under the condition that we get a good number of true positives (i.e., \qG\ pairs whose edit distance is at most $2$), and we do not have too many false positives, it seems that $m = 1.5 \times q$ is a good choice and we set it as the default parameter.
\fi

When $q = 14$, $m = 1.5 \times q$ and $\eta = 1$ (i.e., no filter), we can detect $17.4$ times true positives while introduce $192.3$ times false positives (of the number of exact matches). By setting the filter threshold $\eta = 10^{-5}$, we can detect $12.3$ times true positives while only introduce $17.7$ times false positives. This convinces us that removing frequent \sqGs\ have a greater impact on reducing false positives than true positives, and is thus very useful for our purpose (i.e., to save the verification time at a minimal cost on the accuracy).

\subsection{Finding Overlapping Sequencing Reads}
\label{sec:exp-compare}

In this section we present the experimental results on detecting overlapping sequencing reads with Algorithm~\ref{alg:overlap}.  

\paragraph{Accuracy} We study the precision, recall and $F_1$ scores of all tested algorithms.  The results are presented in Table~\ref{tab:ovlap1} and Table~\ref{tab:ovlap2}. 

\begin{table*}[t]
\centering
  \begin{tabular}{|c|c|c|c|c|c|c|c|c|c|}
    \hline
    \multirow{3}{*}{} &
      \multicolumn{3}{c|}{$\ecoli$} &
      \multicolumn{3}{c|}{$\scere$} &
      \multicolumn{3}{c|}{$\human$} \\

& $Recall$  &$Precision$ & $F_1\ score$  & $Recall$  &$Precision$ & $F_1\ score$ & $Recall$  &$Precision$ & $F_1\ score$ \\
    \hline
    $\mhap$ &78.0\% & 99.8\% & 0.87 &75.1\% & 92.5\% &0.83 &74.1\% &83.7\%  &0.79 \\
    \hline
     $\miniD$ & 92.4\% & 100.0\% &0.96 & 15.4\% & 99.9\% &0.26 & 68.9\% & 98.8\% &0.81 \\
    \hline
     $\miniA$ & 94.1\% & 99.8\% &0.97 & 89.5\% & 93.1\% &0.91 & 71.4\% & 40.4\% &0.52 \\
    \hline
     $\dalign$ & 86.1\% & 97.8\% &0.92 & 82.8\% & 94.8\% &0.88 & 80.6\% & 67.1\% &0.73 \\
    \hline
     $\our$ &95.1\% & 100.0\% &0.97 &90.9\% & 99.2\% &0.95 &84.7\% & 99.2\% &0.91 \\
    \hline
  \end{tabular}
  \caption{Accuracy for pairs with overlaps of lengths $\Gamma \ge 2000$}
\label{tab:ovlap1}
\end{table*}

\begin{table*}[t]
\centering
  \begin{tabular}{|c|c|c|c|c|c|c|c|c|c|}
    \hline
    \multirow{3}{*}{} &
      \multicolumn{3}{c|}{$\ecoli$} &
      \multicolumn{3}{c|}{$\scere$} &
      \multicolumn{3}{c|}{$\human$} \\

& $Recall$  &$Precision$ & $F_1\ score$  & $Recall$  &$Precision$ & $F_1\ score$ & $Recall$  &$Precision$ & $F_1\ score$ \\
    \hline
    $\mhap$ & 66.3\% & 99.8\% &0.80 &65.8\% & 94.3\% &0.77 &77.1\% &84.8\% &0.81 \\
    \hline
     $\miniD$ & 77.2\% & 99.9\% &0.78 & 12.0\% &99.8\% &0.21 & 50.0\% &99.2\% &0.66\\
    \hline
     $\miniA$ & 79.8\% & 99.8\% &0.89 & 72.4\% & 99.6\% &0.84 & 58.2\% & 57.7\% &0.58 \\
    \hline
     $\dalign$ & 79.7\% & 94.3\% &0.86 & 71.8\% &90.9\% &0.80 & 61.5\% &63.7\% &0.63\\
    \hline
     $\our$ & 89.9\% &100.0\% &0.95 &85.1\% & 98.6\% &0.91 &84.7\% &95.5\% &0.90  \\
    \hline
  \end{tabular}
  \caption{Accuracy for pairs with overlaps of lengths $\Gamma \ge 500$}
\label{tab:ovlap2}
\end{table*}

Based on our results, \our\ has the best recall values at {\em all} times, the best precision values in most cases, and the best $F_1$ scores (the harmonic average of precision and recall) at {\em all} times. Its $F_1$ scores are always greater than $0.9$. While the lower bound of the $F_1$ score of the best competitor is only $0.77$ (\mini\ on \scere).
The performance of \our\ is also robust on data from different species and different overlap lengths $\Gamma$.  

We note again that we can further improve the accuracy of \our\ by using multiple embeddings and subsamplings, at the cost of larger space and time.

Comparing the results for the three species, we found that \ecoli\ is generally easier to deal with than \scere\ and \human, which may be due to the fact that \scere\ and \human\ genome contain more repeats.  For different overlap lengths $\Gamma$, we notice that all algorithms generally perform better on the greater length $\Gamma$ than the smaller one, which is reasonable because longer overlaps are generally easier to be detected.

\begin{table*}[t]
\centering
  \begin{tabular}{|c|c|c|c|c|c|c|}
    \hline
    \multirow{3}{*}{} &
      \multicolumn{2}{c|}{$\ecoli$} &
      \multicolumn{2}{c|}{$\scere$} &
      \multicolumn{2}{c|}{$\human$} \\

& $CPU\ Time(s)$  &$Memory(Gb)$ & $CPU\ Time(s)$  &$Memory(Gb)$ & $CPU\ Time(s)$  &$Memory(Gb)$ \\
    \hline
    $\mhap$ &9476  &68.1 &8025 & 75.2  &7472  & 74.6\\
    \hline
     $\miniD$ & 103 & 6.4 & 61 &4.8 &56 & 4.7 \\
    \hline
     $\miniA$ &666  &14.9 &2836 &15.5 &2550 &13.3 \\
    \hline
     $\dalign$ & 2072 & 19.7 & 6376 &17.4 &8821 & 17.6 \\
    \hline
     $\our$ & 6734 & 85.5 &7400 &63.5 &6736 &63.1  \\
    \hline
  \end{tabular}
  \caption{Running time and memory usage}
\label{tab:time}
\end{table*}

\paragraph{Time and Space}  Finally, we study the running time and memory usage of tested algorithms. Our results are presented in Table~\ref{tab:time}.  We observe that \mini\ has the best time and memory performance among all algorithms. \dalign\ spends similar running time as \our, but smaller amount of memory. \our\ has the similar (slightly better) memory and time performance than \mhap.  The reason why \our\  uses relatively large time and memory is that \our\ considers \sqG\ instead of \qG, which captures more matching information between sequences,  and thus needs more time to verify candidate sequence pairs and uses more space. On the other hand, this is also why \our\ significantly improved the accuracy for overlap detection. 

We note that in our experimental studies, we mainly focused on accuracy which we think is the most important; our codes were not fully optimized for space and running time.

\subsection{Summary}
\label{sec:exp-summary}

In this section we have performed an extensive experimental study on \sqG\ and its application to overlap detection.  We observed that the \sqG\ based approach achieved much better accuracy than the conventional \qG\ based approaches for overlap detection, which due to the fact that \sqG\ is capable of capturing near-matches between subsequences.  Employing \sqG\ may introduce a larger number of false positives, but the number can be greatly reduced by applying a frequency-based filter.  The performance of our algorithm is stable and robust on genome sequences from various species that we have tested, and using different overlap lengths $\Gamma$.

\bibliographystyle{ACM-Reference-Format}
\bibliography{paper} 



\end{document}